\documentclass[nojss]{jss}

\usepackage{amstext,amsfonts,amsmath,amssymb,bm,thumbpdf,lmodern,hyperref,dsfont}
\usepackage[all]{hypcap}
\usepackage{enumitem}
\usepackage{color,soul}

\definecolor{changes}{rgb}{1,1,0.7}

\title{The Power of Unbiased Recursive Partitioning:\\
A Unifying View of CTree, MOB, and GUIDE}
\Shorttitle{The Power of Unbiased Recursive Partitioning}

\author{Lisa Schlosser\\Universit\"at Innsbruck
\And Torsten Hothorn\\Universit\"at Z\"urich
\And Achim Zeileis\\Universit\"at Innsbruck}
\Plainauthor{Lisa Schlosser, Torsten Hothorn, Achim Zeileis}

\Abstract{
A core step of every algorithm for learning regression trees is the selection 
of the best splitting variable from the available covariates and the corresponding 
split point. Early tree algorithms (e.g., AID, CART) employed greedy search strategies, 
directly comparing all possible split points in all available covariates. However, 
subsequent research showed that this is biased towards selecting covariates with 
more potential split points. Therefore, unbiased recursive partitioning algorithms 
have been suggested (e.g., QUEST, GUIDE, CTree, MOB) that first select the covariate 
based on statistical inference using $p$-values that are adjusted for the possible split 
points. In a second step a split point optimizing some objective function is selected in 
the chosen split variable. However, different unbiased tree algorithms obtain these 
$p$-values from different inference frameworks and their relative advantages or 
disadvantages are not well understood, yet.

Therefore, three different popular approaches are considered here: classical categorical 
association tests (as in GUIDE), conditional inference (as in CTree), and parameter 
instability tests (as in MOB). First, these are embedded into a common inference framework
encompassing parametric model trees, in particular linear model trees. Second, it is 
assessed how different building blocks from this common framework affect the power of the 
algorithms to select the appropriate covariates for splitting: observation-wise
goodness-of-fit measure (residuals vs.\ model scores), dichotomization of residuals/scores
at zero, and binning of possible split variables. This shows that specifically the 
goodness-of-fit measure is crucial for the power of the procedures, with model scores 
without dichotomization performing much better in many scenarios.
}
\Keywords{classification and regression trees, independence tests, recursive partitioning,
simulation}

\Address{
Lisa Schlosser, Achim Zeileis \\
Universit\"at Innsbruck \\
Department of Statistics \\
Faculty of Economics and Statistics \\
Universit\"atsstr.~15 \\
6020 Innsbruck, Austria \\
E-mail: \email{Lisa.Schlosser@uibk.ac.at}, \email{Achim.Zeileis@R-project.org} \\
URL: \url{https://www.uibk.ac.at/statistics/personal/schlosser-lisa/},\\
\phantom{URL: }\url{https://eeecon.uibk.ac.at/~zeileis/}\\

Torsten Hothorn\\
Universit\"at Z\"urich \\
Institut f\"ur Epidemiologie, Biostatistik und Pr\"avention \\
Hirschengraben 84\\
8001 Z\"urich, Switzerland \\
E-mail: \email{Torsten.Hothorn@R-project.org}\\
URL: \url{http://user.math.uzh.ch/hothorn/}\\

}

\begin{document}

\vspace*{0.4cm}

\section{Introduction}\label{sec:introduction}
In many situations fitting one global model to a given data set can be very challenging, 
especially if the data contains lots of different features with strong variation 
and complex interactions. Therefore, separating the data into more homogeneous subgroups 
based on a set of covariates first can simplify the task and fitting a local model to 
each of the resulting subgroups often leads to better results. This separation can be 
done by applying a tree algorithm. While almost all algorithms proposed in the literature
follow the general idea of splitting the data such that some objective function is optimized
locally, they differ in their specific approaches to selecting a split variable and the 
corresponding split point. 
Some of the first tree algorithms (e.g., AID, \citealp{Morgan+Sonquist:1963}; 
CART, \citealp{Breiman+Friedman+Stone:1984}) rely on exhaustive search procedures to find both, 
the best split point and split variable in one step by directly comparing all possible split
points in all possible split variables. However, it has been shown that this is not only
computationally expensive but also biased towards split variables with many possible split
points \citep{Doyle:1973, Kim+Loh:2001}. 
Therefore, selecting a split variable in a first step and then searching for the 
best split point only within this variable in a separated second step is a more 
promising strategy as applied for example by the algorithms QUEST \citep{Loh+Shih:1997},
%% CHAID \citep{Kass:1980},
GUIDE \citep{Loh:2002}, CTree \citep{Hothorn+Hornik+Zeileis:2006} 
and MOB \citep{Zeileis+Hothorn+Hornik:2008}.
For the first step of selecting a split variable they all share the same basic concept of 
choosing the covariate which shows the highest association to the response variable 
based on $p$-values provided by a statistical test. While QUEST
%%, CHAID,
and GUIDE employ 
statistical significance tests for contingency tables, CTree applies permutation tests in 
a conditional inference framework and MOB uses fluctuation tests based on central limit theorems
for the parameter estimators. 
All these approaches have been shown to work well for various situations, however, the relative
(dis)advantages of the testing strategies have not yet been investigated and compared in detail.

Therefore, in this paper the focus will be put on that first step of tree algorithms,
i.e., the task of selecting the best split variable in a given (sub)sample. In particular, the approach of the 
GUIDE algorithm is compared to the one of the CTree algorithm and the MOB algorithm
by investigating the building blocks of their testing strategies in which they differ:
(1)~Variation of the goodness-of-fit measure for the response: using residuals or full model scores.
(2)~Dichotomization of these residuals or scores.
(3)~Categorization of possible split variables.
Apart from these three main factors further aspects such as the approximation of the null
distribution (conditional vs.\ unconditional) or the type of test statistic (maximally 
selected vs.\ sum of squares) will be considered as well.
For this purpose, a unifying framework for testing strategies in unbiased model-based tree 
algorithms is presented such that each of the three strategies in GUIDE, CTree, and MOB can be 
obtained by a specific combination of the available building blocks. This allows to systematically
vary the building blocks and assess the power of the resulting inference procedure. Moreover,
it is investigated whether the performance of the inference impacts the performance of the
trees differently under pre- vs.\ post-pruning.

In many of the considered scenarios the choice of goodness-of-fit measure heavily influences
the performance of testing strategies. In particular, using model scores leads to
overall clearly better results than employing residuals only. Moreover, the original
values of the goodness-of-fit measure are preferred over dichotomized versions of them.
Also regarding the effects of categorizing possible split variables and the selection
of a pruning strategy clear recommendations can be given based on the presented results.

The remainder of this manuscript is structured as follows:
Section~\ref{sec:urp} reviews unbiased tree models, starting with the general algorithmic idea
(Section~\ref{sec:generic_alg}) followed by the specific algorithms CTree (Section~\ref{sec:ctree}), 
MOB (\ref{sec:mob}), and GUIDE (\ref{sec:guide}), before possible pruning strategies 
(\ref{sec:pruning}) are discussed.
In Section~\ref{sec:unify} the unifying framework for testing strategies in unbiased model-based
recursive partitioning algorithms is presented.
The setting for the simulation study is introduced in Section~\ref{sec:simulation}
and the results are illustrated and discussed in Section~\ref{sec:results}.

\section{Unbiased recursive partitioning}
\label{sec:urp}

\subsection{Generic algorithm}
\label{sec:generic_alg}
The basic idea of building a regression tree model is to partition the data into smaller 
and more homogeneous subgroups based on a set of covariates. Various tree 
algorithms have been developed, following essentially the same general structure,
employing the covariates as \emph{split variables} in the tree induction.
Starting at the root of the tree, pertaining to the full available data sample, the
algorithms proceed in the following steps:
\begin{enumerate}
\item For the current sample a (possibly simple) model is fitted by optimizing
some objective function (or loss function) that reflects the goodness of fit.
\item Among all available split variables one is selected as the split variable,
choosing the split point such that goodness of fit is maximized in the
resulting subgroups.
\item Steps 1 and 2 are repeated within each subgroup until some stopping
criterion is attained.
\end{enumerate}
The term ``model'' is used here in a very broad sense and encompasses not
only least-squares or maximum-likelihood models but also simple constant
fits such as means or average proportions.
Therefore, depending on the type of model, the employed objective function can 
for example be the sum of deviations from a typical/average value, but it can 
also be based on a model for a response along with potential split variables.
For instance, different types of residuals (or the signs thereof) can be employed
\citep[see e.g.,][]{Loh:2002} as well as rank sums or logrank scores
\citep[as in][]{Hothorn+Hornik+VanDeWiel:2006}.

As explained in Section~\ref{sec:introduction} the considered tree algorithms CTree, 
MOB, and GUIDE first select the split variable and then, in a separate step, the 
split point. For the first step of selecting a split variable they all apply statistical 
tests following the same basic strategy: 
\begin{enumerate}
\item To capture how the objective function changes with the observations in
the current subgroup, a disaggregated, observation-wise goodness-of-fit
measure is obtained. More formally, this is an $N \times K$ matrix
where $N \in \mathbb{N}$ is the number of observations and $K \in \mathbb{N}$ 
the number of goodness-of-fit measures per observation.
\item The dependency or association of this goodness-of-fit matrix with
each possible split variable $Z_j$, $j \in \{1, \dots, J\}$, is assessed
using some suitable test statistic. The corresponding $p$-values allow for
a comparison of all $J$ split variables on a standardized or unified scale and in
that way for an unbiased split variable selection.
\item The split variable corresponding to the smallest $p$-value -- and thus the
highest influence on the goodness of fit of the model -- is selected for 
splitting the data into subgroups.
\end{enumerate}
As an example -- and explained in more detail below -- consider a linear regression tree.
Thus, a linear regression model is fitted in each subgroup, minimizing the
residual sum of squares as the aggregated goodness-of-fit measure. The corresponding
observation-wise goodness-of-fit measure can be given by the residuals or the scores
(gradient contributions). Analogously, the log-likelihood and corresponding score function 
could be used.

While this basic approach is the same for the GUIDE, CTree, and MOB algorithms, they differ 
in their strategies on how to calculate test statistics. In order to point out these specific 
characteristics in Section~\ref{sec:unify} the strategies of the three tree algorithms are first 
explained in more detail in Sections~\ref{sec:ctree}, \ref{sec:mob}, and \ref{sec:guide}.

\subsection{CTree}\label{sec:ctree}

The CTree algorithm \citep{Hothorn+Hornik+Zeileis:2006} is based on the idea of providing 
non-parametric regression tree models in a conditional inference framework by applying 
permutation tests. 
To select a split variable it is tested whether there is any association between the transformed 
response $h(Y)$ and each possible transformed split variable $g(Z_j)$, $j = 1, \dots, J$.
The only requirement for the function $h$ is to depend on $Y$ in a permutation-symmetric way but
this encompasses ranks, scores, indicator functions, etc.\ and can also be multidimensional.
For a numeric response the identity function $h(Y) = Y$ is a common 
choice while a categorical response can be mapped to a unity vector by an indicator function 
$h(Y) = (0,\ldots,1,\ldots,0)^{\top}$.
Alternatively, the function $h$ can capture location and scale of $Y$ via
$h(Y) = (Y, (Y-\bar{Y})^2)^{\top}$. 
If a parametric model is fitted to the response $Y$ with some covariate(s)~$X$ employed as 
\textit{regressor} variable(s),
then a model-based transformation $h(Y) = s(Y,X,\hat{\beta})$ can be used, e.g.,
the residuals in a linear model with regression coefficients $\beta$. Moreover,
$s$ can be the score function pertaining to the objective function (or loss function)
$\ell$:
$$
s(Y, X, \beta) = \frac{\partial \ell(Y, X, \beta)}{\partial \beta}
$$
The estimate of the model parameters $\hat{\beta}$ is obtained by optimizing
the sum of the objective function $\ell$, aggregated over all observations.
This framework includes many different M-type estimators as special cases, including maximum 
likelihood and ordinary least squares estimation. For a $K$-dimensional parameter
$\beta$ the score function evaluated for the $i$-th observation $s(y_i, x_i, \beta)$
is also a $K$-dimensional vector, i.e., the gradient contribution of the $i$-th observation.
Thus, the $N \times K$-matrix consisting of these scores or gradient contributions for all 
$i = 1, \dots, N$ observations is a natural candidate for the observation-wise goodness-of-fit
measure as described in the previous Section~\ref{sec:generic_alg}.

Similarly, different types of functions can be chosen for the influence function $g$ depending
on a possible split variable $Z$. A simple choice in case of $Z$ being a numeric variable is again
the identity function $g(Z) = Z$. For categorical variables $g$ can also map its values to the 
corresponding unity vectors by an indicator function $g(Z) = (0,\ldots,1,\ldots,0)^{\top}$.

To test for independence of $h(Y)$ and $g(Z)$ CTree calculates a linear association test statistic,
following the framework of \cite{Strasser+Weber:1999}. The corresponding conditional expectation and
covariance given all permutations of the response variable can be calculated and used to 
standardize the test statistic. This standardized statistic has an asymptotic normal distribution
which is in fact multivariate if either of the transformation $h(Y)$ and/or $g(Z)$ is
multivariate. The actual test is carried out by mapping this standardized statistic to
the real line either by taking the absolute maximum or using a quadratic form -- with $p$-values
being computed from the analogous transformation of the normal distribution
\citep[see][for a hands-on introduction and Appendix~\ref{app:teststat_ctree} for more details on the linear test statistic]{Hothorn+Hornik+VanDeWiel:2006}.

If both variables $Y$ and $Z$ are numeric the default independence test corresponds to a Pearson 
correlation test. For one numeric and one categorical variable essentially a one-way ANOVA
(analysis of variance) is employed while for two categorical variables a $\chi^2$~test is performed.
Thus, in the general CTree framework many types of tests can be specified by selecting suitable
transformations $g$ and $h$. While originally
conceived for nonparametric models, it is easy to adapt CTree to model-based testing and 
recursive partitioning by choosing a model-based $h$ transformation as argued above
\citep[see the concrete examples in][]{Zeileis+Hothorn:2013,Seibold+Zeileis+Hothorn:2016}.

\subsection{MOB}\label{sec:mob}

In contrast to CTree the MOB algorithm \citep{Zeileis+Hothorn+Hornik:2008} was explicitly
designed for a model-based goodness-of-fit measure in order to embed parametric models
into a regression tree framework. Thus, MOB is based on an objective/loss function $\ell$
and corresponding score function $s$. The original paper considered generalized linear models
(GLMs) and survival regression models but subsequently various other models have been applied as well,
including beta regression \citep{Gruen+Kosmidis+Zeileis:2012}, psychometric item response theory
models \citep{Strobl+Kopf+Zeileis:2015}, or mixed effects models \citep{Fokkema+Smits+Zeileis:2018}.
But just like CTree can be applied to parametric models, MOB conversely also encompasses
simple regression and classification trees, e.g., by choosing an intercept-only model.

For selecting a split variable MOB employs a score-based test that relies on the central limit
theorem for the parameter estimate $\hat{\beta}$. The test assesses whether the scores
-- when ordered by the potential split variable $Z$ -- fluctuate randomly around their
zero mean or differ systematically in certain subgroups. The latter would indicate a
parameter instability that could be captured by fitting separate models (optimizing $\ell$)
in the resulting subgroups. In case of a numeric split variable $Z$ both the score-based statistic
and the partitioned objective function $\ell$ are maximized over all possible splits in $Z$
(subject to certain minimal subgroup size constraints). Unlike the CTree framework, MOB
relies on classical unconditional inference. For more details see \cite{Zeileis+Hornik:2007} and 
Appendix~\ref{app:teststat_mob}.

\subsection{GUIDE}\label{sec:guide}

Building on earlier work for the QUEST algorithm \citep{Loh+Shih:1997}, \cite{Loh:2002}
proposed the GUIDE algorithm blending trees with parametric regression models and
encompassing simpler classification and regression trees as special cases. Thus,
linear regression models could be fitted in the nodes of a tree as well as
constant fits such as simple mean response~\mbox{$\hat{\beta} = \bar{y} = \frac{1}{N}\sum_{i=1}^N y_i$}.
The tests for selecting the split variable are then based on the corresponding residuals,
e.g., for a simple linear regression \citep[as used in][]{Loh:2002}:
$$
r(Y,X,\hat{\beta}) = Y - \hat{\beta_0} - \hat{\beta_1} \cdot X.
$$
In subsequent work other models together with an appropriate choice of residuals
have been applied such as in
regression trees for longitudinal and multiresponse data \citep{Loh+Zheng:2013}, 
%(http://www.stat.rice.edu/~jrojo/4th-Lehmann/slides/Loh.pdf)
quantile regression models \citep{Chaudhuri+Loh:2002}, 
and proportional hazards modeling via Poisson regression \citep{Loh+He+Man:2015}, among others.

To construct a statistical test two additional transformations are carried out:
(1)~The residuals $r(Y,X,\hat{\beta})$ are dichotomized at zero, yielding an indicator for positive vs.\ negative residuals.
(2)~Each possible split variable $Z$ is categorized, i.e., unless $Z$ is already categorical
it is split at its quartiles into four bins.
Subsequently, a $\chi^2$~test of independence is performed for the dichotomized residuals and
each categorized/categorical split variable.
After choosing the split variable showing the highest dependency by yielding the lowest 
$p$-value, the split point minimizing the overall goodness-of-fit measure is selected. 
Note that the split point in numeric variables $Z$ is searched over all possible splits,
not just the four bins that were constructed for the $\chi^2$~test. More details on the
applied test statistic can be found in Appendix~\ref{app:teststat_guide}.

%\newpage
\subsection{Pruning}
\label{sec:pruning}

To avoid overfitting recursive partitioning algorithms need to assure that trees do not grow
too large. Apart from certain minimal subgroup size or maximal tree depth constraints this is
classically accomplished by so-called ``pruning'' approaches. As the three tree algorithms
considered here (CTree, MOB, and GUIDE) differ in their default choice of pruning approach,
we briefly discuss these here. However, as all three algorithms can in principle be combined
with any of the pruning approaches this is done only relatively briefly.

The classical CART algorithm \citep{Breiman+Friedman+Stone:1984} proposed to first grow a large
tree and then prune those splits in the tree that did not increase predictive performance in a
cross-validation. This is also know as \emph{post-pruning} (after growing the initial tree) and
more specifically cost-complexity pruning.

In the unbiased recursive partitioning literature this post-pruning approach is also used frequently
\citep[e.g., in][]{Loh+Vanichsetakul:1988,Loh+Shih:1997,Kim+Loh:2001,Loh:2002} and the $p$-values
from the association tests are only employed for selecting the split variable on a unified scale.
However, \cite{Hothorn+Hornik+Zeileis:2006} proposed to also use these $p$-values for a 
so-called \emph{pre-pruning} strategy which stops growing the tree as soon as no significant association
can be found in a given subgroup. This approach is the default in CTree and also in
MOB. However, \cite{Zeileis+Hothorn+Hornik:2008} also pointed out that a natural strategy for
post-pruning in model-based partitioning is to use information criteria such as AIC (Akaike information
criterion) or BIC (Bayes information criterion), following the ideas of \cite{Su+Wang+Fan:2004}.

Clearly, for inference-based pre-pruning it is crucial that the association tests employed for
split selection work well as statistical significance test, i.e., conform with their nominal size
and have high power. In contrast, when using post-pruning (either based on cross-validation or
information criteria) it might not be as crucial that the significance test works well and has
high power.

Due to these considerations we first evaluate the significance tests underlying CTree, MOB,
and GUIDE by themselves, i.e., without growing an actual tree in combination with a pruning strategy.
Subsequently we combine the tests with a cost-complexity post-pruning approach in order to assess
whether shortcomings of the tests are mitigated by pruning.

\section{Unifying framework}\label{sec:unify}

Each of the algorithms CTree, GUIDE, and MOB can be characterized by its combination
of the type of model fits, tests, and pruning strategy employed to
grow the tree. Table~\ref{tab:combinations} provides an overview of the default
combinations.
However, as discussed above, subsequent publications have emphasized that all three
algorithms can be combined with different model fitting approaches and to some degree
different pruning strategies have been explored as well. Thus, the class of tests
employed for the unbiased splitting variable selection forms the core of each of the
algorithms: conditional inference (CTree) vs.\ score-based fluctuation tests (MOB)
vs.\ residual-based $\chi^2$ tests (GUIDE). Therefore, we consider a standard class
of model fits (namely, linear regression trees) and investigate the relative
advantages and disadvantages of the tests themselves (which are essential to pre-pruning)
as well as the combination with post-pruning. Thus, subsequently
the names CTree, MOB, and GUIDE distinguish the test-based variable and split selection
rather than the entire algorithm with all default settings.

\begin{table}[t!]
\centering
\begin{tabular}{ l l l l }
\hline\noalign{\smallskip} 
& Fit             & Test                     & Pruning \\ 
\noalign{\smallskip}\hline\noalign{\smallskip}
CTree & Non-parametric  & Conditional inference    & Pre\\
MOB   & Parametric      & Score-based fluctuation  & Pre (or post with AIC/BIC)\\
GUIDE & Parametric      & Residual-based $\chi^2$  & Post (cost-complexity pruning)\\
\noalign{\smallskip}\hline
\end{tabular}
\caption{Default combinations of fitted model type, test type and pruning strategy
for the algorithms CTree, MOB, and GUIDE.
\label{tab:combinations}}
\end{table}

\subsection{Building blocks of testing strategies}\label{sec:buildingblocks}

Even though CTree, MOB, and GUIDE differ in the specific tests they apply, their approaches for split
variable selection follow the same basic structure as explained in Section~\ref{sec:generic_alg}. 
In fact, the tests can be embedded in a unifying conceptual framework that yields the
different tests by combining various building blocks.
These mostly differ in the way the model for the dependent variable on the one hand and
the splitting variables on the other are prepared or transformed:
\begin{itemize}
\item \emph{Goodness-of-fit measure:}\\
Different variations of the disaggregated, observation-wise goodness-of-fit measure of the 
model for the response $Y$ and possible regressors $X$ can be considered. Either residuals
$r(Y,X,\hat{\beta})$ can be used as proposed for GUIDE or model scores $s(Y,X,\hat{\beta})$
as proposed for MOB. All three algorithms can, in principle, use both goodness-of-fit
measures though, which is probably brought out most clearly in the CTree algorithm that
explicitly allowed for different transformations $h(Y)$ in its original description already.
\item \emph{Dichotomization of residuals/scores:}\\
Rather than testing independence between the split variables and the residuals/scores
themselves, it is possible	
to dichotomize residuals/scores at $0$ so that only their signs are assessed (as proposed for GUIDE).
\item \emph{Categorization of split variables:}\\
Similarly, the split variables can also be categorized (for testing only). This was proposed for
GUIDE, employing binning at the quartiles yielding four categories of approximately
equal size.
\end{itemize}
The three algorithms combine these building blocks in different ways as shown in Table~\ref{tab:buildingblocks}:
When applying CTree for unbiased model-based recursive partitioning it has been suggested to use
the model scores without dichotomization and assess their association with the untransformed split
variables using a conditional inference test. This is similar to a squared correlation test statistic.
MOB also employs the scores without
dichotomization and maximally selects a score statistic over all potential split points in the
split variable. GUIDE employs the dichotomized residuals and assesses their association with
the categorized split variable in a classical (unconditional) $\chi^2$ test.

But the building blocks could be easily re-combined to yield new types of tests. For example,
in the GUIDE approach, a one-way ANOVA can be used for assessing the association of the residuals
(without dichotomization) with the categorized split variable. Or alternatively, a multivariate
one-way ANOVA can be used for the non-dichotomized scores as opposed to the residuals etc.

Note that there are further differences in the testing strategies between the three algorithms, e.g.,
using conditional vs.\ unconditional approximations of the null distributions. However, this difference
has relatively little influence compared to the other building blocks considered in detail here.
Moreover, both similarities and relative differences between these approaches have been previously
discussed, e.g., in \cite{Hothorn+Zeileis:2008} and \cite{Zeileis+Hothorn:2013}.

\begin{table}[t!]
\centering
\begin{tabular}{ l l l l l }
\hline\noalign{\smallskip} 
& Scores       & Dich.  & Cat.  & Statistic\\
\noalign{\smallskip}\hline\noalign{\smallskip}
CTree & Model scores & --	     & -- 	       & Sum of squares\\
MOB   & Model scores & --	     & -- 	       & Maximally selected\\
GUIDE & Residuals    & \checkmark    & \checkmark      & Sum of squares\\
\noalign{\smallskip}\hline
\end{tabular}
\caption{Testing strategies of CTree, MOB, and GUIDE with the corresponding
setting of the building blocks in the unifying framework and the type of test statistic.
\label{tab:buildingblocks}}
\end{table}

%\newpage

\subsection{Linear model tree}\label{sec:lmtrees}

To focus on the unified testing framework, as described in the previous section, we employ
the same model fits for all three algorithms. To do so, we employ linear regression models
because it is such a basic and widely used model and linear model trees were the leading
illustrations in both the original MOB \citep{Zeileis+Hothorn+Hornik:2008} and GUIDE \citep{Loh:2002} papers.
However, the conclusions drawn from this example also hold for many other model types.

To fix notation, we consider the following models for the simulation 
study in Section~\ref{sec:simulation}:
$$
Y = \beta_0 + \beta_1 \cdot X + \epsilon
$$
with response variable $Y$, regressor variable $X$, and error term $\epsilon$.
In particular, in the investigated tree models the coefficients $\beta_0$ and $\beta_1$
can depend on the possible split variables $Z_j$, $j=1,\ldots,J$, such that
$$
Y = \beta_0(Z_1, \ldots, Z_J) + \beta_1(Z_1, \ldots, Z_J) \cdot X + \epsilon.
$$
This model is fitted, as usual, by ordinary least squares (OLS) to the observations in each notation,
yielding the parameter estimates $\hat{\beta} = (\hat{\beta_0}, \hat{\beta_1})^\top$.
To keep notation simple we present the following equations for the root node with all observations
$\{(y_i,x_i)\}_{i=1,\ldots,N}$ and $N \in \mathbb{N}$. In subsequent nodes the same equations
are used but just for a smaller subgroup.

The aggregated goodness-of-fit measure (or objective or loss function) in OLS estimation
is the sum of squared residuals: $\sum_{i=1}^N \ell(y_i,x_i,\beta_0,\beta_1)$ where
$$
\ell(y_i,x_i,\beta_0,\beta_1) = r(y_i,x_i,\beta_0,\beta_1)^2
$$
is the squared residual which is defined as
$$
r(y_i,x_i,\beta_0,\beta_1) = y_i - \beta_0 - \beta_1 \cdot x_i.
$$
These residuals can also intuitively be used as the corresponding disaggregated
observation-wise goodness-of-fit measure. Another natural candidate for this is the
score function for the $i$-th observation:
$$
s(y_i, x_i, \beta_0, \beta_1) = 
\frac{\partial \ell}{\partial (\beta_0, \beta_1)^{\top}}(y_i, x_i, \beta_0, \beta_1) = 
-2 \cdot r(y_i,x_i, \beta_0, \beta_1) \cdot (1,x_i)^\top
$$
Thus, up to a constant scaling factor of $-2$ (that could also be omitted) the first 
component of the scores is in fact the residual. 
However, it is complemented by a second component
that captures the slope effect of $x_i$. Therefore, when computing the score for all $N$ observations
this yields an $N \times 2$ score matrix whose first column corresponds to the $N$ residuals.
As this is the derivative with respect to the intercept parameter~$\beta_0$, it
captures changes in the intercept. Moreover, the second column of the score matrix contains
derivatives with respect to the slope parameter~$\beta_1$ and thus captures changes in this.

Hence, tests based on the full scores (as in CTree and MOB) include residual-based tests (as 
in GUIDE) as a special case. Therefore, score-based tests can in principle capture all changes 
in the objective function that residual-based tests can capture -- but the reverse is not
necessarily true. In the next sections we will investigate how relevant this is in practice 
and how much it depends on the concrete test statistics employed.

\section{Simulation setting and evaluation}
\label{sec:simulation}

In this simulation study two different scenarios are considered for the linear model trees
presented in Section~\ref{sec:lmtrees}. First, the underlying tree structure based on which the
data is generated is a stump, i.e., a tree with only one split (``stump'' scenario, see 
Figure~\ref{fig:dgp_stump}). 
By keeping the tree so simple we can focus on the testing strategy only giving focus to their
power in terms of selecting the correct split variable.

In the second scenario the true tree structure contains two splits in two different variables
yielding a tree with three terminal nodes (``tree'' scenario, see Figure~\ref{fig:dgp_tree}). 
It employs the same basic structure as the first scenario but simply adds another split. This 
allows to evaluate the power of the three testing strategies in a more complex setting, in 
combination with using a post-pruning strategy.

\subsection{Data generating process}

\subsubsection{``Stump'' scenario}
Each generated data set consists of the response and regressor variables, one true split 
variable and nine noise split variables as listed in Table~\ref{tab:variables} 
together with the corresponding distributions.

\begin{table}[t!]
\centering
\begin{tabular}{l l l}
\hline\noalign{\smallskip} 
Name                    & Notation                   & Specification \\
\noalign{\smallskip}\hline\noalign{\smallskip} 
\textit{Variables:}     &                        & \\ 
Response                & $Y$                    & $=\beta_0(Z_1) + \beta_1(Z_1) \cdot X + \epsilon$ \\
Regressor               & $X$                    & $\mathcal{U}([-1,1])$ \\
Error                   & $\epsilon$             & $\mathcal{N}(0,1)$ \\
True split variable     & $Z_1$                  & $\mathcal{U}([-1,1])$\\
Noise split variables   & $Z_2$--$Z_{10}$   & $\mathcal{U}([-1,1])$  or $\mathcal{N}(0, 1)$ \\
                        &                        & (alternating) \\
\noalign{\smallskip}\hline\noalign{\smallskip} 
\multicolumn{3}{l}{\textit{Parameters/functions:}} \\ 
Intercept                      & $\beta_0$  & $0$ or $\pm \delta$\\
Slope                          & $\beta_1$  & $1$ or $\pm \delta$\\
True split point               & $\xi$      & $\in \{0, 0.2, 0.5, 0.8\}$ \\
Effect size                    & $\delta$   & $\in \{0, 0.1, 0.2, \ldots , 1\}$ \\
\noalign{\smallskip}\hline 
\end{tabular}
\caption{\label{tab:variables}Variables included in the data generating process 
as used for the ``stump'' scenario. In the ``tree'' scenario $Z_2$ is also a true split variable, 
not only $Z_1$, and hence $\beta_0$ and $\beta_1$ are functions $\beta_{k-1}(Z_1,Z_2)$, $k=1,2=K$.}
\end{table}

\begin{figure}[p!]
\setkeys{Gin}{width=1\linewidth}
\minipage{0.45\textwidth}
\includegraphics{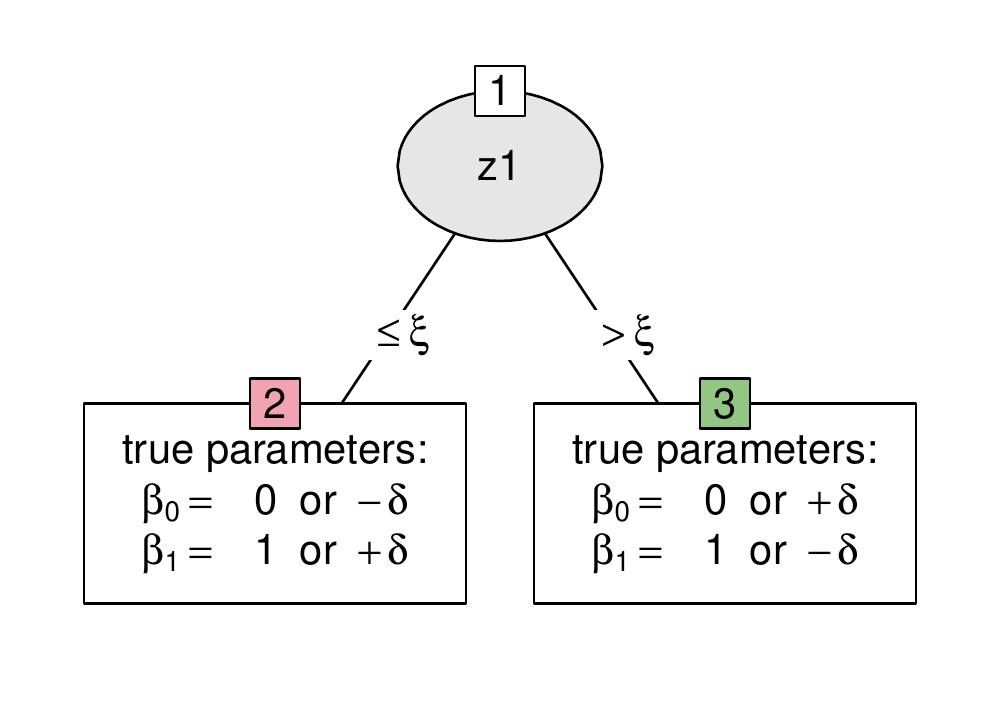}
\endminipage
\minipage{0.45\textwidth}
\includegraphics{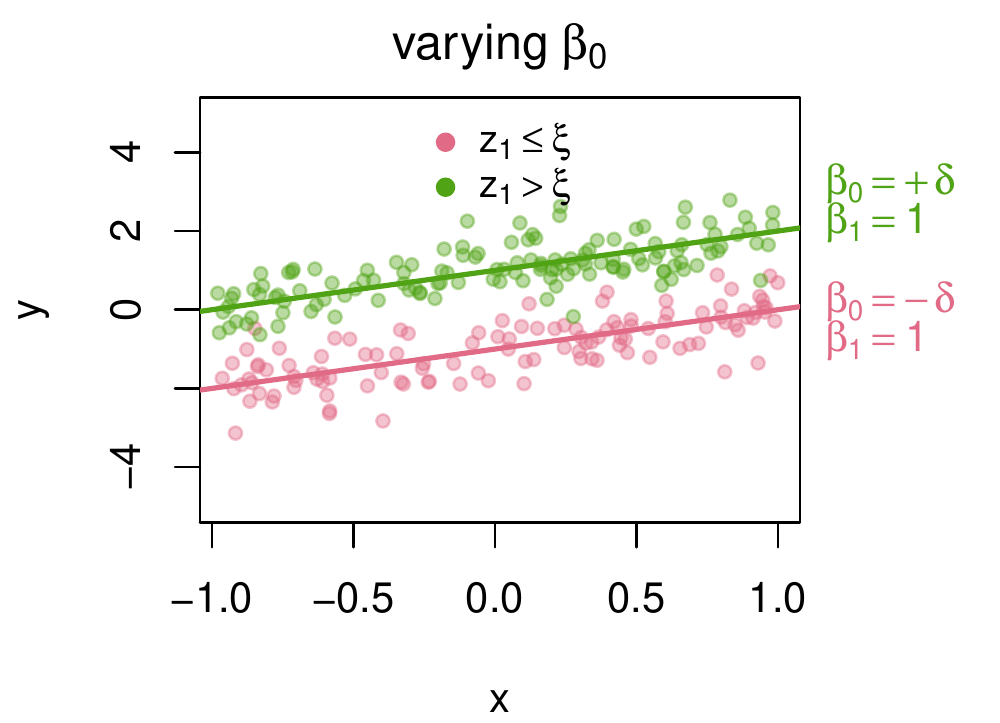}
\endminipage

\minipage{0.45\textwidth}
\includegraphics{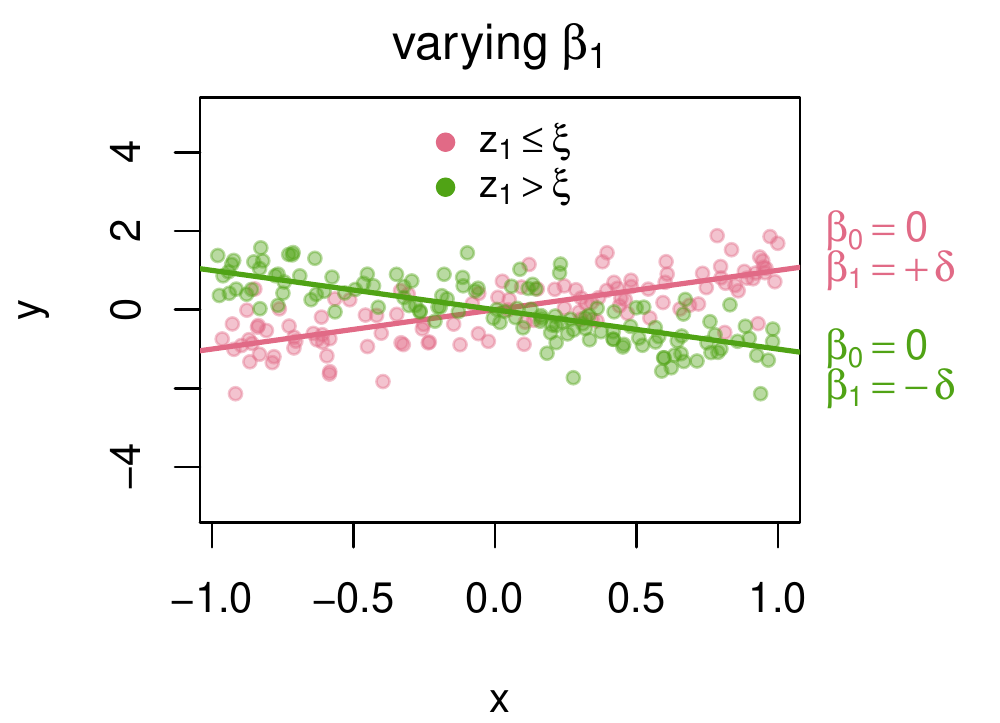}
\endminipage
\minipage{0.45\textwidth}
\includegraphics{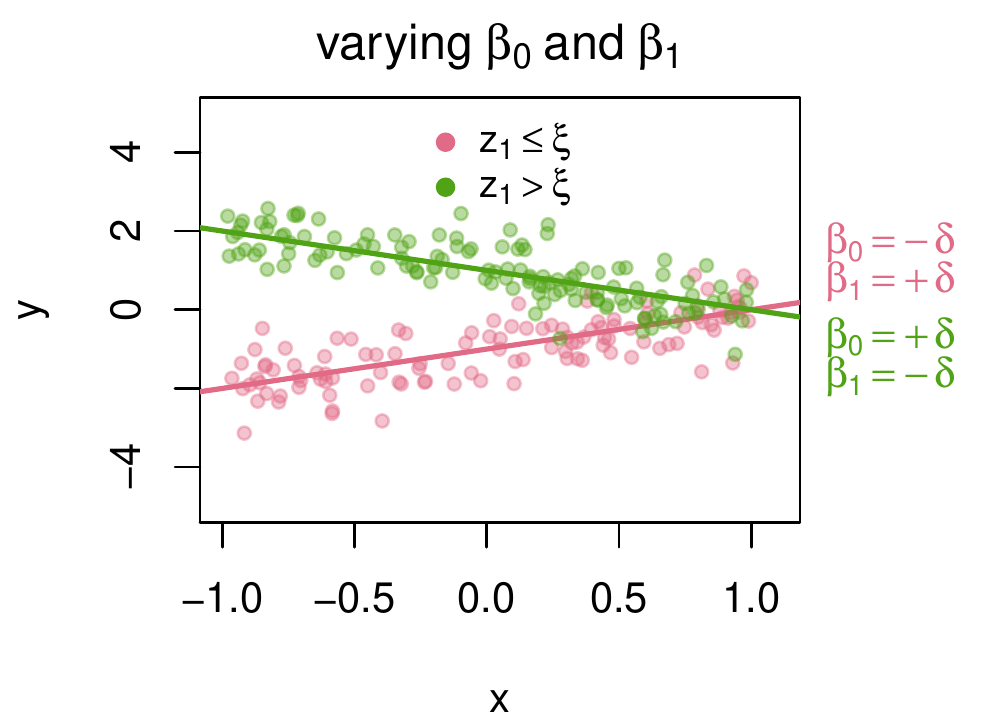}
\endminipage
\caption{\label{fig:dgp_stump}{Top left panel: True stump structure applied in the data generating process for the ``stump'' scenario with the true model parameters $\beta_0$
and $\beta_1$, either varying or being fixed at $0$ or $1$ respectively.
Top right and bottom panels: Bivariate plot of the response $Y$ on regressor $X$ 
with effect size $\delta = 1$ and for three variations: 
varying $\beta_0$ and fixed $\beta_1$ (top right),
fixed $\beta_0$ and varying $\beta_1$ (bottom left),
varying $\beta_0$ and varying $\beta_1$ (bottom right).}}
\end{figure}

The location parameter $\mu$ of the normally distributed response variable $Y$ depends
linearly on the regressor variable $X$.
% which is a uniformly distributed variable taking values between $-1$ and $1$.
% can either be a binary variable (taking either $0$ or $1$) or a uniformly distributed 
% variable (taking values between $-1$ and $1$).
Moreover, the intercept $\beta_0$ and/or the slope parameter $\beta_1$
can depend on the true split variable $Z_1$. More specifically, three different variations
are considered for the coefficients $\beta_0$ and $\beta_1$:
\begin{enumerate}
\item The intercept $\beta_0$ varies depending on $Z_1$ while $\beta_1$ is fixed (at $1$).
\item The slope coefficient $\beta_1$ varies depending on $Z_1$ while $\beta_0$ is fixed 
(at $0$).
\item Both coefficients $\beta_0$ and $\beta_1$ vary depending on $Z_1$.
\end{enumerate}
See Figure~\ref{fig:dgp_stump} for an illustration. More precisely, the coefficient that
changes~($\beta_{k-1}$ with $k=1,2=K$) switches between two values at the split point $\xi$:
\begin{align*}
\beta_{k-1}(Z_1) = \begin{cases}
-\delta \cdot (-1)^{k-1}  \quad \text{if } Z_1 \leq \xi \\
+\delta \cdot (-1)^{k-1}  \quad \text{if } Z_1 > \xi
\end{cases}
\end{align*}
In that way, the type of variation is the same for $\beta_0$ and $\beta_1$, however, in 
opposite directions.

\begin{figure}[p!]
\begin{center}
\setkeys{Gin}{width=1\linewidth}
\minipage{0.5\textwidth}
\includegraphics{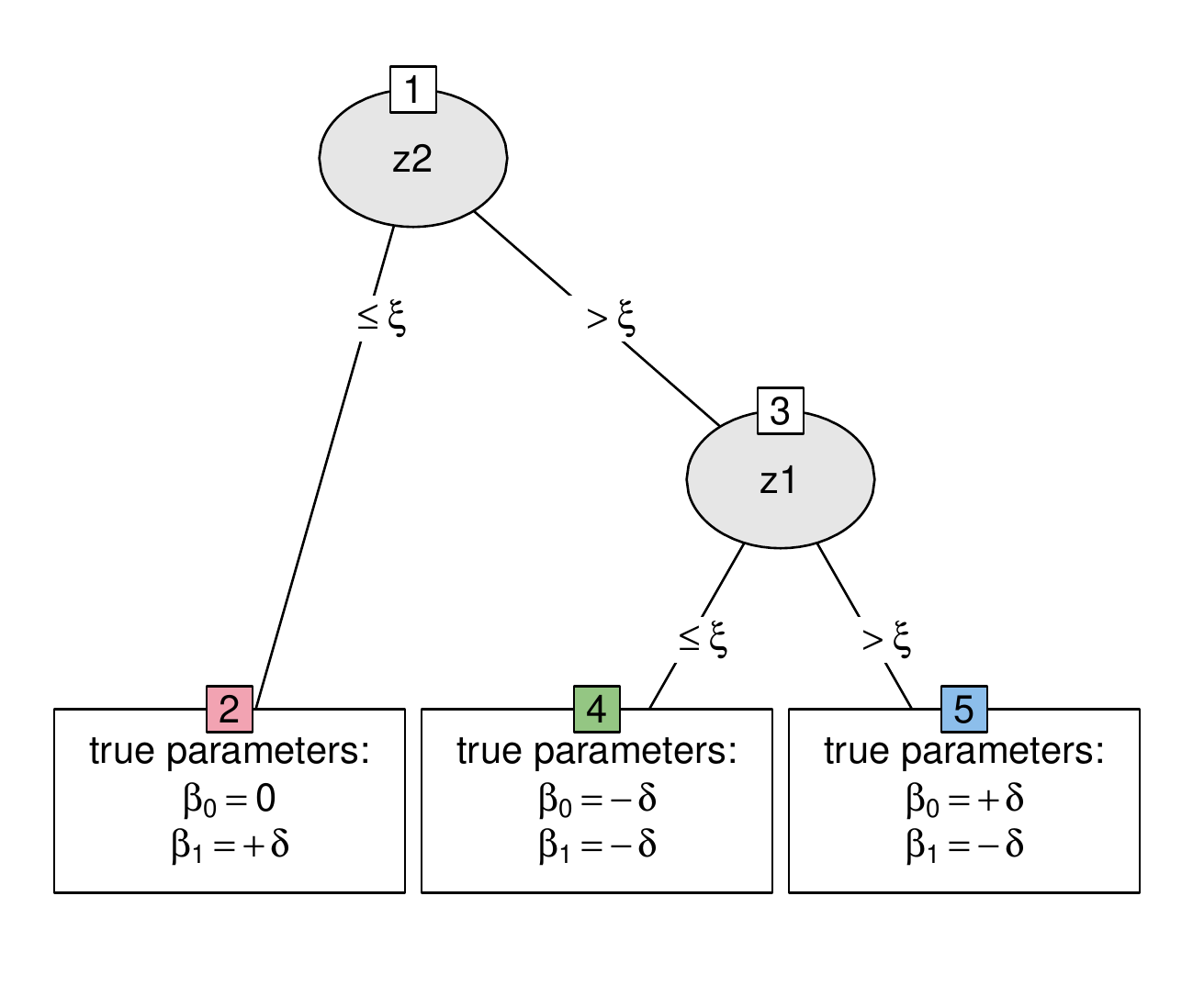}
\endminipage
\minipage{0.45\textwidth}
\includegraphics{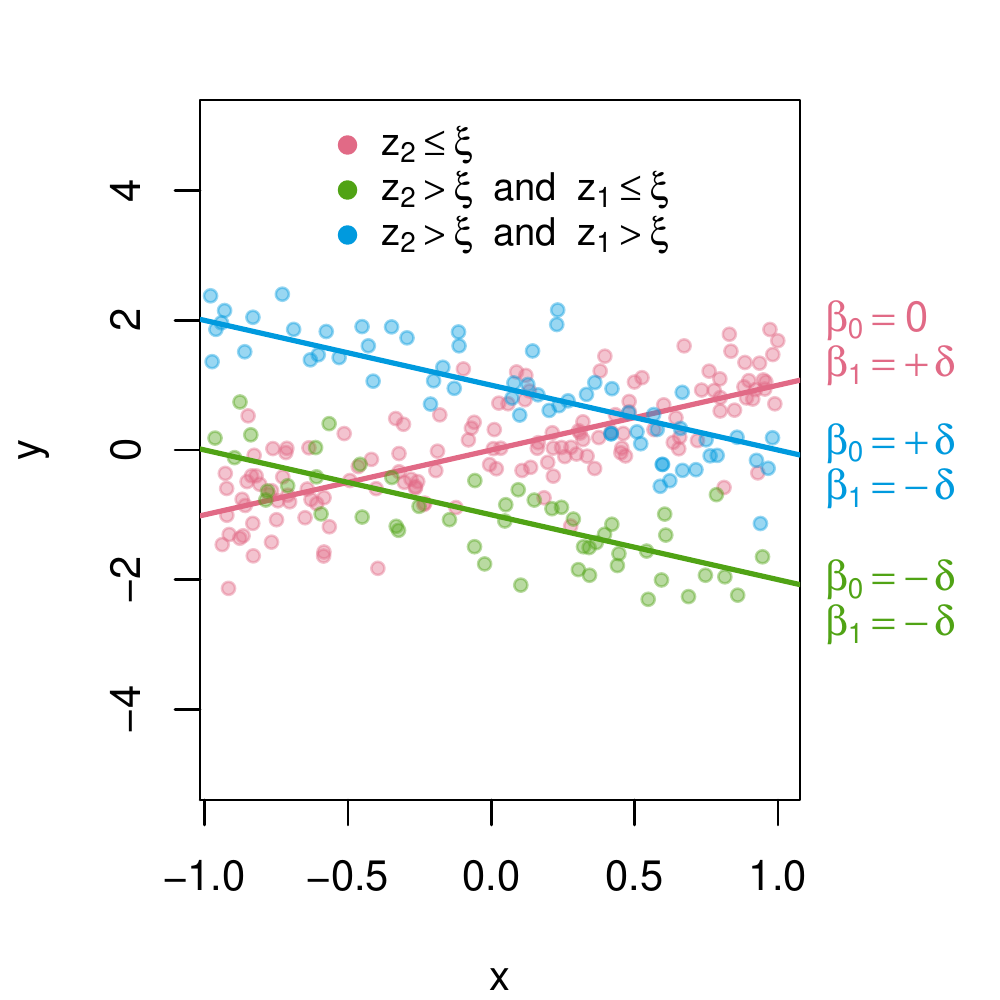}
\endminipage
\caption{\label{fig:dgp_tree}Left panel: True tree structure
applied in the data generating process for the ``tree'' scenario 
with the true model parameters $\beta_0$ and $\beta_1$ for each terminal panel.
Right panel: Bivariate plot of the response $Y$ on regressor $X$ 
with effect size $\delta = 1$.}
\end{center}
\end{figure}

\subsubsection{``Tree'' scenario}

For the second simulation scenario (``tree'' scenario, see Figure~\ref{fig:dgp_tree}) the same basic setting with the same 
variables as in the ``stump'' scenario is applied (see Table~\ref{tab:variables}).
However, not only $Z_1$ but also $Z_2$ is used as a true split variable following a uniform distribution
on $[-1,1]$. Therefore, an additional split
is preformed yielding a tree with three terminal nodes. The first split (in $Z_2$, at~$\xi$) induces a 
change in the slope parameter $\beta_1$ while the second split (in $Z_1$, also at~$\xi$) corresponds 
to a change in the intercept $\beta_0$.
Hence, contrary to the ``stump'' scenario where three variations are considered for the parameters 
$\beta_0$ and $\beta_1$, only this one variation with both parameters varying is investigated for 
the ``tree'' scenario. In particular, the parameters depend on the split variables in the following way:
\begin{align*}
\beta_{0}(Z_1, Z_2) = \begin{cases}
0       \qquad \text{if } Z_2 \leq \xi \\
-\delta  \quad \text{ if } Z_1 \leq \xi \land Z_2 >\xi \\
+\delta  \quad \text{ if } Z_1 > \xi \land Z_2 >\xi
\end{cases}
\end{align*}
\begin{align*}
\beta_{1}(Z_1, Z_2) = \beta_{1}(Z_2)= \begin{cases}
+\delta \quad \text{if } Z_2 \leq \xi \\
-\delta \quad \text{if } Z_2 > \xi
\end{cases}
\end{align*}

\subsection{Evaluation}
\label{sec:evaluation}
The testing strategies are evaluated over a stepwise increasing effect size $\delta$, 
on 100 replications per step each consisting of 250 observations. 
To compare the performance of the evaluated testing strategies for each step 
in the ``stump'' scenario the following criteria are considered:
the $p$-values pertaining only to the true split variable $Z_1$;
and the proportion of replications for which the $p$-value of $Z_1$
is the lowest and significant at 5\% level (i.e., where $Z_1$ would be
selected for splitting in a pre-pruning approach), denoted by the 
``selection probability''. In that way, the power of the
considered statistical tests can be compared as they are all applied
as significance tests answering two questions at once:
(1)~Should a split be performed at all? 
(2)~If so, in which variable?
For the first question the $p$-values regarding all available split variables are compared 
to a predefined level of significance $\alpha = 0.05$. Only if the smallest $p$-value is smaller
than $\alpha$ a split is performed and the split variable corresponding to this $p$-value is selected. 

For the ``tree'' scenario the adjusted Rand index~(ARI) is calculated 
as a measure of similarity between the true tree structure and the fitted model tree.

As explained before, the aim of this simulation study is to investigate the effects
of each particular building block of the unifying framework presented in Section~\ref{sec:unify} 
rather than the whole testing strategies. Therefore, other combinations as presented in 
Table~\ref{tab:buildingblocks}, hence adapted versions of GUIDE, CTREE, and MOB are evaluated as well
as their original versions. 

\section{Results}
\label{sec:results}

In the following we first investigate the properties of the testing
strategies in the ``stump'' scenario, focusing on the tests' $p$-values
for the true split variable $Z_1$ and its corresponding selection
probability (i.e., the association of $Y$ and $Z_1$ being significant at 5\% level 
and having the lowest $p$-value among all split variables). Section~\ref{sec:scores} 
begins by highlighting the importance of using full model scores vs.\ residuals only
before Section~\ref{sec:3way} considers all building blocks (scores vs.\ residuals,
dichotomization of these, and categorization of the split variable).
Subsequently, the ``tree'' scenario is employed to investigate how the
performance of the tests affects growing the trees overall. This is
evaluated using the adjusted Rand index for trees grown by pre-pruning and 
cost-complexity post-pruning (Section~\ref{sec:tree}).

% To illustrate the effects of the building blocks in the unifying framework presented in 
% Section~\ref{sec:unify} different combinations of them are investigated for the first
% simulation scenario (stump) in Sections~\ref{sec:scores} and~\ref{sec:3way}. 
% Then the testing strategies are evaluated on the more
% complex tree structure of the second scenario in Section~\ref{sec:tree}. 
% The corresponding results are shown in the following plots.

\newpage
\subsection[``Stump'' scenario: Residuals vs. full model scores]{``Stump'' scenario: Residuals vs.\ full model scores}
\label{sec:scores}

A crucial difference between the testing strategies of CTree/MOB and GUIDE
is the difference between using only the residuals vs.\ the full model
scores. In the literature on structural change tests it is well-established
that residual-based tests can only capture parameter differences that
affect the conditional mean \citep[see e.g.,][ and further discussion in 
Section~\ref{sec:discussion}]{Ploberger+Kraemer:1992}.
Hence we compare the CTree, MOB, and GUIDE algorithms  -- all three using the 
default specification as shown in Table~\ref{tab:buildingblocks} -- and 
additionally consider a new GUIDE flavor, denoted GUIDE+scores, that uses
dichotomized scores rather than residuals.

\begin{figure}[t!]
\setkeys{Gin}{width=\linewidth}
\includegraphics{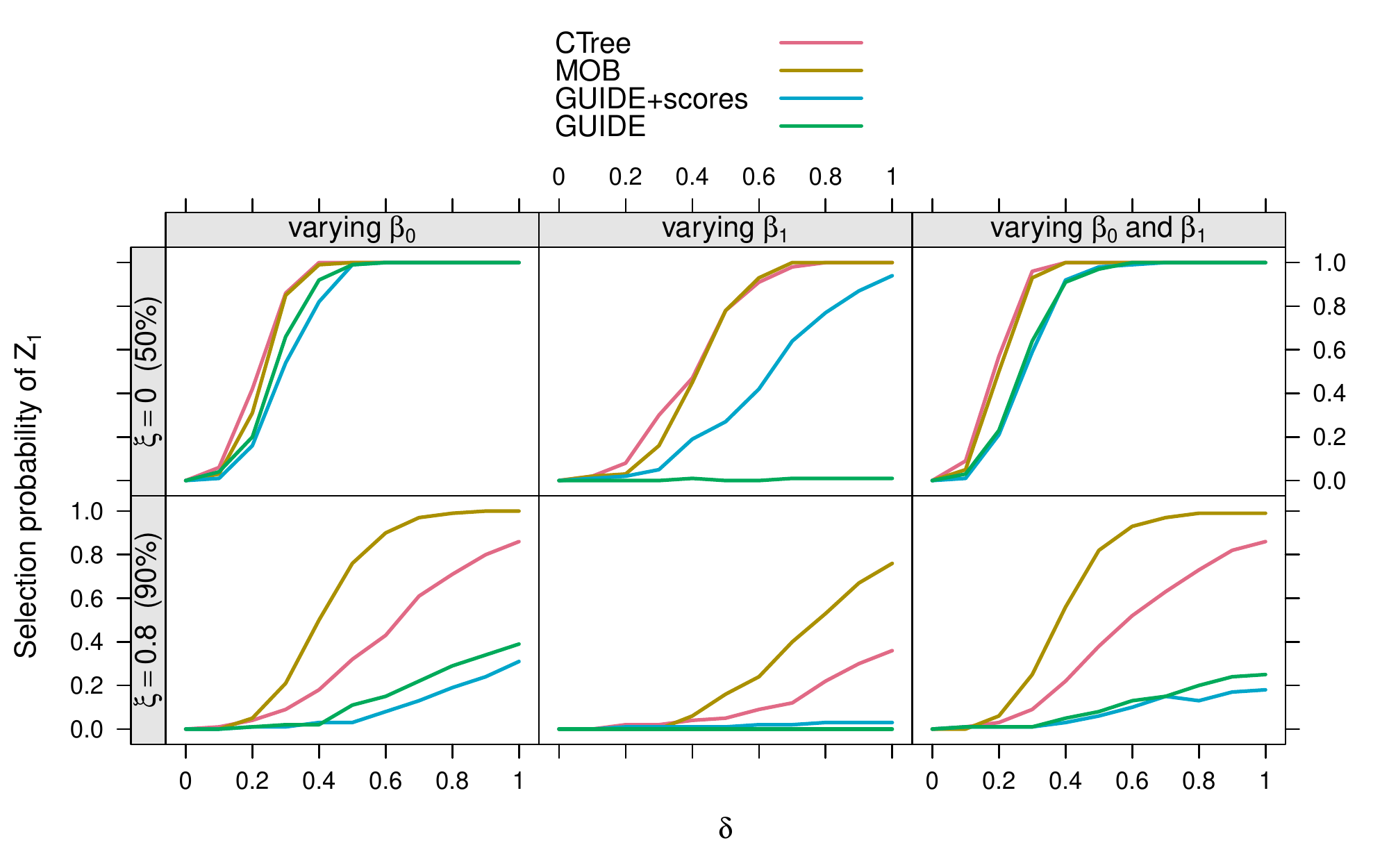}
\caption{\label{fig:scores_stump}
Selection probability of the true split variable $Z_1$ for
testing strategies CTree, MOB, GUIDE, and GUIDE+scores in the ``stump''
scenario. Probabilities are estimated over increasing effect size $\delta$
with 100 replications of 250 observations per step. The true split point
in $Z_1$ is either the median 0 (top) or the 90\% quantile 0.8 (bottom)
for either varying intercept $\beta_0$ (left), varying slope $\beta_1$
(middle), or both coefficients varying (right).}
\end{figure}

In Figure~\ref{fig:scores_stump} the performances of the testing strategies are 
represented by the corresponding selection probability, i.e., a significance level 
is incorporated. (Appendix~\ref{app:significance} shows that the same qualitative
conclusions can be drawn when the significance level is \emph{not} included.)
For a split point at the median ($\xi=0$, top row) all testing strategies 
perform similarly well as long as the intercept varies (left and right panel) with CTree and 
MOB being only slightly ahead. However, for the split point at the $90$\% quantile $\xi = 0.8$ 
(bottom row) the performance of all of the applied strategies decreases. 
While both GUIDE versions struggle to detect the correct split variable even for a high effect 
size, MOB is clearly ahead with CTree leading to the second best results.
This advantage of MOB over CTree is mainly due to the abrupt shift in the 
model parameters $\beta_0$ and $\beta_1$ and turns into an advantage of CTree over MOB for 
a smooth transition with continuously-changing parameters (see Appendix~\ref{app:contbeta}).

However, in the scenario where only $\beta_1$ (but not the intercept $\beta_0$) is affected by the split,
the residual-based GUIDE approach has no power at all even for a true split at the median (top middle panel).
It is easily possible, though, to substantially mitigate this problem by using scores
(sensitive to changes in all parameters) rather than residuals only (sensitive to changes
in the conditional mean). The remaining difference between MOB/CTree and GUIDE+scores
is due to dichotomizing the scores at zero and due to categorizing the split variables
which are investigated in more detail in the following section.

\subsection{``Stump'' scenario: Full factorial analysis of building blocks}\label{sec:3way}

To investigate the impact of each of the building blocks separately the most general 
case of the ``stump'' scenario where the intercept and slope parameter are both varying is 
considered. In this evaluation all possible combinations of the building blocks 
have been included. The different levels of each of the three building blocks are listed in 
Table~\ref{tab:levels} where $h$ and $g$ refer to the transformation functions applied to the 
response $Y$ or a split variable $Z$ respectively, both as described in Section~\ref{sec:ctree}. 

In the case of categorization, the split variable $Z$ is binned at the
quartiles. This corresponds to a four-dimensional 0/1 transformation
function $g(Z)$ that indicates into which of the bins each observation falls.
Maximum selection across potential split points in a variable $Z$ also
corresponds to a multivariate 0/1 transformation function $g(Z)$. However,
in this case for each potential split point an indicator is used that is 0
before and 1 after the respective split point.
(See also Table~\ref{tab:combinations_appendix} in Appendix~\ref{app:combinations} 
for a more detailed overview of all 12 combinations of the building blocks.)

\begin{table}
\centering
\begin{tabular}{ l l l }
\hline\noalign{\smallskip} 
Building block               & Levels              & Transformation \\
\noalign{\smallskip}\hline\noalign{\smallskip} 
Residuals vs.\ scores        & \textit{residuals}  & $h(Y) = r(Y,X,\hat{\beta})$\\
($h$ transformation)         & \textit{scores}     & $h(Y) = s(Y,X,\hat{\beta})$\\
\noalign{\smallskip}\hline\noalign{\smallskip} 
Dichotomization              & \textit{yes}        & $\mathds{1}_{[0,\infty)}(h)$\\
(of $h$)                     & \textit{no}         & $h$ without\\
                             &                     & \quad dichotomization\\
\noalign{\smallskip}\hline\noalign{\smallskip} 
Categorization               & \textit{cat}        & $g(Z)=(0,1,0,0)^{\top}$\\
($g$ transformation)         &                     & \quad indicating the\\
                             &                     & \quad assigned bin\\
& \textit{max}               & $g(Z)=(0,\ldots,0,1,\ldots,1)^{\top}$\\
&                            & \quad indicating the\\
&                            & \quad potential split point\\
& \textit{lin}        & $g(Z)=Z$\\
\noalign{\smallskip}\hline
\end{tabular}
\caption{Available levels of the building blocks.
\label{tab:levels}}
\end{table}

\begin{figure}[t!]
\setkeys{Gin}{width=1\linewidth}
\includegraphics{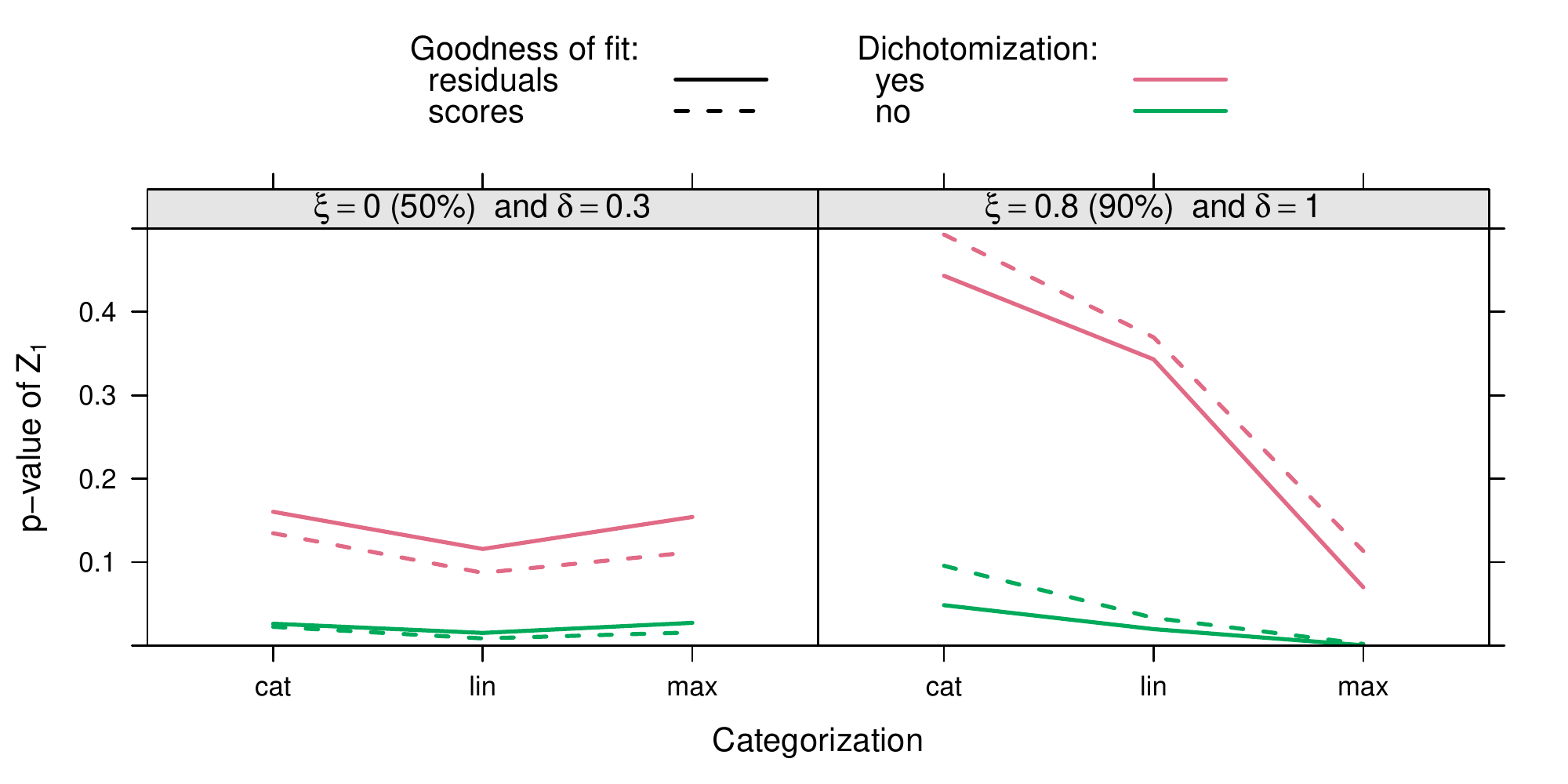}
\caption{\label{fig:3way} Effects of the three building blocks \textit{residuals vs.\ scores},
\textit{dichotomization}, and \textit{categorization} in the ``stump'' scenario with the intercept
and the slope parameter both varying. All possible combinations have been 
evaluated in two different settings and their performances are compared based on the 
mean $p$-values corresponding to the true split variable $Z_1$. 
Left panel: true split point at the median $0$ and with
effect size $0.3$.
Right panel: true split point at the $90$\% quantile $0.8$ and with effect 
size $1$.}
\end{figure}

Based on the results displayed in Figure~\ref{fig:3way} it can be stated that dichotomizing 
residuals/scores decreases the performance as it leads to higher $p$-values for
the true split variable $Z_1$ and thus to lower power in the settings considered. 
For a true split at the median ($\xi=0$, left panel) this effect is almost constant
across the three types of categorization considered. However, in case 
of the true split point at the $90$\% quantile $\xi = 0.8$ (right panel), categorizing split 
variables increases the $p$-value of $Z_1$ even more. 
Not surprisingly, maximum selection is most advantageous in this case (i.e., for a late abrupt shift)
which is harder to detect based on a linear statistic. But overall it depends on the situation
whether a linear or a maximum selection of a split variable leads to lower $p$-values. 

Comparing the two panels suggests that a categorization weakens the performance of the
tests unless the true split point is close to one of the breaks from the binning
(as in the left panel).
Moreover, while the effect of both transformations (dichotomization of residuals/scores and
categorization of split variables) can be observed separately in Figure~\ref{fig:3way},
combining them increases the negative impact remarkably.

As already shown in Section~\ref{sec:scores} the use of scores vs.\ residuals has a minor 
effect if there is a change in the intercept which is
supported by the small differences between the dashed lines (scores) and the solid
lines (residuals) in Figure~\ref{fig:3way}.

Note that the effect size $\delta$ in the results from Figure~\ref{fig:scores_stump} has been chosen
so that $p$-values for the non-dichotomized tests are roughly comparable: $\delta = 0.3$
for the true split point at the median $\xi = 0$ (left) vs.\ a stronger effect of $\delta = 1$
for the true split point at the 90\% quantile $\xi = 0.8$ (right).
Additional evaluations for varying effect size $\delta$ can be found in Appendix~\ref{app:increasingdelta}.

\subsection[``Tree'' scenario: Pre-pruning vs. post-pruning]{``Tree'' scenario: Pre-pruning vs.\ post-pruning}
\label{sec:tree}

So far the testing strategies underlying the different tree algorithms have only been considered
as classical significance tests, i.e., in terms of power and $p$-values. However, one could argue
that for a tree this is practically not really relevant -- at least when combined with a post-pruning
strategy such as cost-complexity pruning \citep{Breiman+Friedman+Stone:1984}. In the latter case it only matters that the relevant split
variables have the lowest $p$-value among all potential split variables -- but it is irrelevant whether
this is significant or not. To investigate to which extent this is actually true we evaluate the
different tree algorithms in the more complex ``tree'' scenario (see Figure~\ref{fig:dgp_tree}):
once with significance-based pre-pruning and once with cost-complexity post-pruning.

Recall that the true split structure is composed of splits in two different variables, both at
the same split point $\xi$. First, the split in $Z_2$ changes the slope from $+\delta$ to
$-\delta$. Second, the split in $Z_1$ changes the intercept in the negative slope group from
$-\delta$ to $+\delta$. In the simulation the effect size $\delta$ is increased from 0 to 1
for different split points $\xi$ from 0 to 0.8. Here, the performance is not evaluated in terms
of test properties but only in terms of tree properties, namely the adjusted Rand index (ARI)
in comparison to the true partition of the data.

\begin{figure}[t!]
\setkeys{Gin}{width=\linewidth}
\includegraphics{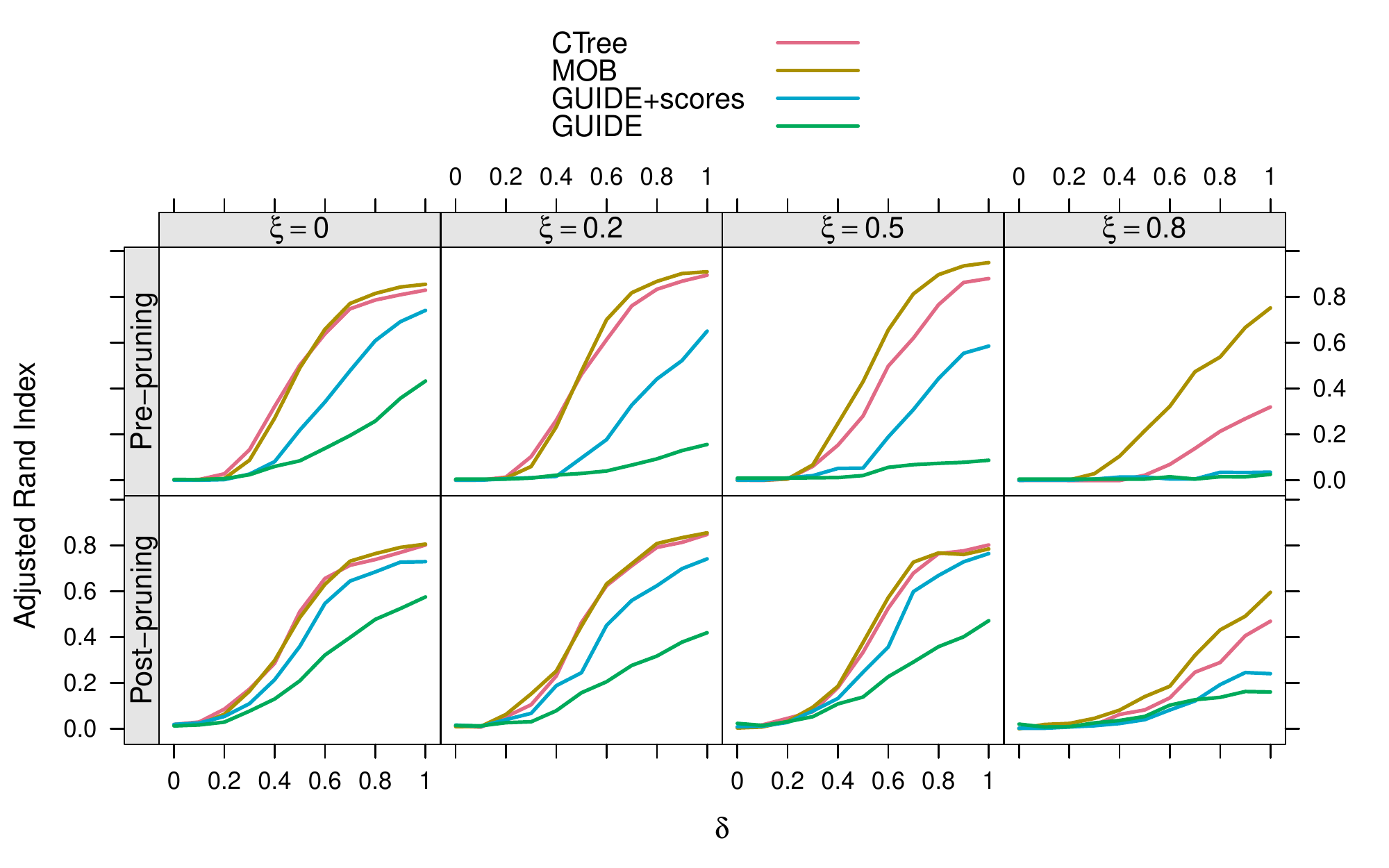}
\caption{\label{fig:tree_all}
Adjusted Rand index (ARI) for the testing strategies CTree, MOB, GUIDE, and GUIDE+scores in the
``tree'' scenario, once applying pre-pruning (top row) and once post-pruning (bottom row). 
The ARI is evaluated over
increasing effect size $\delta$ with 100 replications of 250 observations per step.
Each panel represents a different value of the true split point $\xi$ ($0$, $0.2$, $0.5$ and 
$0.8$ from left to right, corresponding to the $50$\%, $60$\%, $75$\%, and $90$\% quantile).
}
\end{figure}

The top row of Figure~\ref{fig:tree_all} presents the results for pre-pruning which, not surprisingly,
reflect the results from previous sections. Thus, CTree and MOB perform similarly and rather well
as the underlying significance tests have good power properties. Only for late abrupt shifts the
linear test statistics in CTree clearly have lower power than the maximally-selected statistics in MOB.
Again, this picture would reverse for smooth changes instead of abrupt shifts. Compared to CTree/MOB
both GUIDE flavors clearly perform worse as the underlying tests are less powerful; notably
for the residual-based GUIDE which again has problems picking up the slope change associated with $Z_2$.

When switching from pre-pruning to post-pruning (bottom row of Figure~\ref{fig:tree_all})
it is indeed shown that many of the problems stemming from the low power of the two GUIDE
flavors are indeed mitigated. Thus, by first growing a large tree without stopping upon
non-significance and then pruning back based on predictive performance instead substantially
improves the fit of the two GUIDE flavors. However, using the residuals only in GUIDE still
performs clearly worse compared to the GUIDE+scores. The latter is essentially on par with
CTree and MOB when the true split matches one of the categorization bins (i.e., $\xi = 0$
and $\xi = 0.5$, respectively) while CTree/MOB still perform somewhat better in the other
cases ($\xi = 0.2$ and $\xi = 0.8$).

In summary, there is clear support for the conventional wisdom that the power of the testing
strategy in an unbiased tree algorithm is not so important when combined with post-pruning.
However, there are limits to this. Consequently, when tests can be made more powerful --
e.g., by using scores instead of residuals -- then this improves the corresponding tree
algorithm. Finally, the simulation also supports using pre-pruning in an unbiased tree algorithm
when the underlying testing strategy also works well as a classical significance test.

\section{Discussion}
\label{sec:discussion}

The testing strategies underlying the unbiased recursive partitioning algorithms
CTree, MOB, and GUIDE have been embedded in a common inference framework, highlighting
what the tests have in common and what sets them apart. Concerning the effects 
of the corresponding building blocks for the tests, three main conclusions can be drawn: 
\begin{itemize}
\item \emph{Goodness-of-fit measure:} Assessing all dimensions of a model via the full
scores is to be preferred over assessing only a subset with the residuals. The former
can substantially improve performance while leading only to minor deteriorations
when it is not necessary.
\item \emph{Dichotomization of residuals/scores:} No scenarios could be found where this
is beneficial and tests without dichotomization performed clearly better in several
scenarios.
\item \emph{Categorization of split variables:} The effects of categorization are not so
clear-cut and depend on the true data structure. If there are indeed abrupt shifts
close to the breaks from the categorization, it works well. However, for splits
close to the margins performance can deteriorate and a maximally-selected test is
preferable. Finally, a linear statistic performs better for smooth rather than
abrupt changes.
\end{itemize}
Thus, also when categorization as in GUIDE is used we would recommend to employ the non-dichotomized
scores instead of the dichotomized residuals. Note that such a test corresponds to a
multiple ANOVAs (for the score components) and is easily available in statistical software.
Moreover, in the \proglang{R} system the \pkg{coin} package provides a convenient toolbox that
encompasses multivariate transformation functions $h$ and/or $g$.

Linear models as presented in this study have been chosen as they are highly relevant in practice,
allow for simple illustrations, and theoretical insights are available for the testing strategies.
However, the results can be easily extended to a wide variety of other models where the introduced building
blocks can be applied in the same way. Specifically, it has been shown theoretically that certain
changes in the parameters do not lead to shifts in the residuals (but in other components of the
scores). \cite{Ploberger+Kraemer:1992} showed that residual-based tests can detect a change in the parameters
of a linear model only if it also causes a shift in the expected value $\mathbb{E}(Y)$. 
This is not the case if changes are orthogonal to the mean regressor which in our case is $(1,0)^{\top}$.
Consequently, if only the slope $\beta_1$ but not the intercept $\beta_0$ changes, the shift
is of type $(0, \delta)^\top$ and thus orthogonal to the mean regressor. Due to this residual-based
tests as in GUIDE break down and do not have power to detect this. Note that this situation does not
have to be rare in practice: Especially for binary regressor variables (e.g., as in treatment-subgroup
investigations) it can easily occur \citep[see Figure~2 in][for an illustration]{Loh+He+Man:2015}.

Also in more general models, residual-based and score-based procedures are expected to perform
equally well if all model parameters are highly correlated. But if parameters do change orthogonally
this might again be missed when only considering residuals -- and full model scores are typically
easily available as the appropriate remedy. Note that the score function~$s$ can also
simply be seen as a transformation of the response variable $Y$ (and potentially regressors $X$)
to a different space in order to allow for a well structured analysis of dependencies. This has
been exploited in several tree-based approaches previously published in the literature,
e.g., in \cite{Hothorn+Zeileis:2017} and \cite{Schlosser+Hothorn+Stauffer:2019}. Similarly, it would
be of interest to investigate a score-based version of the extended GUIDE algorithms beyond the
linear model, e.g., in \citealp{Loh+Zheng:2013}, \citealp{Chaudhuri+Loh:2002}, and \citealp{Loh+He+Man:2015}.

\section*{Computational details}
\label{sec:comp}
The applied implementation is based on the \proglang{R} package \pkg{partykit}
(version~1.2.4) which is available 
on \proglang{R}-Forge at \url{https://R-Forge.R-project.org/projects/partykit/}.
The code to reproduce the simulation study is available in the supplement for this 
paper on arXiv.org E-Print Archive (\url{https://arxiv.org/}).

The functions \code{ctree} and \code{mob} provide an implementation of the
two tree algorithms in their original form. For their adapted versions additional
modifications have been applied within these functions allowing for a categorization
of possible split variables and a dichotomization of scores.
To evaluate the GUIDE algorithm a reimplementation of this algorithm has been 
built using the basic framework of \code{ctree} and \code{mob}.

\section*{Acknowledgments}
Torsten Hothorn received funding from the Swiss National Science
Foundation, grant number 200021\_184603.

\bibliography{ref.bib}

\newpage

\begin{appendix}

\section{Test statistics}\label{app:teststat}
This section provides further information on the test statistics applied in CTree, MOB, 
and GUIDE. Most of the presented details have been extracted from the original papers 
\cite{Hothorn+Hornik+Zeileis:2006}, \cite{Zeileis+Hothorn+Hornik:2008}, 
and \cite{Loh:2002}, however, notation is adapted to the main manuscript
in order to allow for better comparison.

\subsection{CTree}\label{app:teststat_ctree}
To measure the association of a response $Y$ and each possible split variable $Z_j$, $j=1,\ldots,J$,
the CTree algorithm applies a linear test statistic $T_j$ which is excerpted from Section~3.1.
(``Variable selection and stopping criteria'') of the original paper 
\citep{Hothorn+Hornik+Zeileis:2006} and is of the following form:
$$
T_j(Y, Z_j, w) = \text{vec}\left( \sum_{i=1}^N w_i g_j(Z_{ji})h(Y_i)^{\top}\right)
$$
where $w$ are optional weights, $g_j: \mathcal{Z}_j \rightarrow \mathbb{R}^{P}$ is a nonrandom 
transformation of the split variable $Z_j$ and the influence function 
$h: \mathcal{Y} \times \mathcal{Y}^N \rightarrow \mathbb{R}^Q$ depends on the response 
$Y = (Y_1, \ldots, Y_N)$ in a permutation symmetric way and is set to 
$h(Y_i) = s(Y_i, X_i, \hat{\beta})$ for a score-based approach. 
Moreover, by applying the ``vec'' operator the resulting
$P~\times~Q$ matrix is converted into a $PQ$ column vector.
Following \cite{Strasser+Weber:1999}
the conditional expectation $\mu_j$ and covariance $\Sigma_j$ of a linear test statistic
$T_j$ can be calculated and used to standardize an observed linear test statistic $t_j$ within a 
function $c$ mapping into the real line.
For example, the maximum of the absolute values of the standardized linear statistic
$$
c_{\max}(t_j,\mu_j,\Sigma_j) = \max \left| \frac{(t_j-\mu_j)}{\text{diag}(\Sigma_j)} \right|
$$
or a quadratic form
$$
c_{\text{quad}}(t_j, \mu_j, \Sigma_j) = (t_j - \mu_j)\Sigma_j^{+} (t_j - \mu_j)^{\top}
$$
can be considered where $\Sigma_j^{+}$ is the Moore-Penrose inverse of $\Sigma_j$.
Since the asymptotic conditional distribution of a linear test statistic $T_j$ is a 
multivariate normal with parameters $\mu_j$ and $\Sigma_j$ \citep{Strasser+Weber:1999}, 
the asymptotic distribution of $c_{\text{max}}$ is normal while the quadratic form 
$c_{\text{quad}}$ follows an asymptotic $\chi^2$~distribution. Based on this knowledge 
the corresponding $p$-values can be calculated easily.

\subsection{MOB}\label{app:teststat_mob}
The MOB algorithm employs an empirical fluctuation process $W_j$ to measure deviations of
the model scores $s$ from zero with respect to an ordering induced by the possible split variable
$Z_j$, $j = 1,\ldots,J$. As described in detail in Section~3.2. (``Testing for 
parameter instability'') of the original paper \citep{Zeileis+Hothorn+Hornik:2008}
this process is of the following form:
$$
W_j(t) = \hat{V}^{-1/2} N^{-1/2} \sum_{i=1}^{\lfloor Nt \rfloor} s_{\pi(Z_{ji})} \qquad (0 \leq t \leq 1)
$$
with the model scores $s_i = s(Y_i, X_i, \hat{\beta})$ being sorted by the possible split variable $Z_j$ by including the ordering permutation $\pi(Z_{ji})$. 
To scale this partial sum process an estimate $\hat{V}$ of the covariance matrix $\text{cov}(s(Y,X,\hat{\beta}))$ is included.
Following \cite{Zeileis+Hornik:2007} $W_j(t)$ converges to a Brownian bridge $W^0$ under the null
hypothesis of parameter stability. To obtain a test statistic a scalar functional $\lambda(\cdot)$ capturing the fluctuation in the empirical process can be applied and the corresponding asymptotic distribution of $\lambda(W_j(t))$ can be obtained by employing the same functional to the limit process, i.e., $\lambda(W^0)$. 

One possible and intuitive choice for a functional in order to asses instabilities over a numerical 
split variable is the following:
$$\lambda_\text{supLM}(W_j) = \max_{i = \underline{i},\ldots,\overline{i}}\left(\frac{i}{N}\cdot \frac{N-i}{N}\right)^{-1} \left|\left|W_j\left(\frac{i}{N}\right)\right|\right|_2^2
$$
where a minimal segment size $\underline{i}$ and then $\overline{i} = N - \underline{i}$ are used to 
define the interval $[\underline{i},\overline{i}]$. Other possible functionals, for example also
for categorical split variables, and more details,
particularly on calculating the corresponding $p$-values can be found in Section~3.2. of the 
original paper \citep{Zeileis+Hothorn+Hornik:2008}.

\subsection{GUIDE}\label{app:teststat_guide}
The test statistic of the $\chi^2$~test as applied in the GUIDE algorithm is
$$
X^2 = \frac{(O_{11} - E_{11})^2}{E_{11}} + \ldots + \frac{(O_{24} - E_{24})^2}{E_{24}}
$$
where $O_{lm}$ are the observed frequencies of observations in a certain combination of 
dichotomized residuals ($l=1,2$) and categorized split variables ($m = 1,\ldots,4$). $E_{lm}$ 
are the corresponding expected frequencies under independence 
$E_{lm} = \frac{O_{l \cdot} \cdot O_{\cdot m}}{N}$. 
Under the null hypothesis of independence the asymptotic distribution of $X^2$ is a 
$\chi^2$~distribution allowing for a straight-forward calculation of $p$-values.
If full model scores are used instead of residuals a sum of the test statistic over the 
number of distribution parameters is considered such that each summand corresponds to one 
column of the score matrix. For this case the degrees of freedom of the resulting 
$\chi^2$~distribution is then also the corresponding sum of degrees of freedom over the 
number of distribution parameters.

\section{Combinations of building blocks applied}\label{app:combinations}

For the evaluation of all 12 possible combinations of the presented building blocks
the original testing strategies CTree, MOB, and GUIDE have been applied together
with modified versions of them. The employed versions and the corresponding combination
of building blocks are listed in Table~\ref{tab:combinations_appendix}.
%Each suffix indicates a change of the original form of the considered testing strategy.
%In particular, it relates to the transformations of the data before applying a statistical
%test as represented by the different building blocks. In that way,
%``+res'' refers to using only residuals (instead of all model scores),
%``+scores'' to using all model scores (instead of only residuals),
%``+cat'' to applying a categorization of split variables,
%``+bin'' to applying a dichotomization of scores/residuals.

\begin{table}[p!]
\centering
\begin{tabular}{ l l l l l}
\hline\noalign{\smallskip} 
Residuals/Scores   & Dich. & Cat. & \multicolumn{2}{l}{Via testing strategy}\\
\noalign{\smallskip}\hline\noalign{\smallskip}
residuals	   & yes	     & cat            & GUIDE & {(default)}\\
residuals	   & yes	     & max            & MOB   & {(modified)}\\
residuals	   & yes	     & lin            & CTree & {(modified)}\\
residuals	   & no 	     & cat            & MOB   & {(modified)}\\
residuals	   & no 	     & max            & MOB   & {(modified)}\\
residuals	   & no 	     & lin            & CTree & {(modified)}\\
scores  	   & yes	     & cat            & GUIDE & {(modified)}\\
scores  	   & yes	     & max            & MOB   & {(modified)}\\
scores  	   & yes	     & lin            & CTree & {(modified)}\\
scores  	   & no 	     & cat            & MOB   & {(modified)}\\
scores  	   & no 	     & max            & MOB   & {(default)}\\
scores  	   & no 	     & lin            & CTree & {(default)}\\
\noalign{\smallskip}\hline
\end{tabular}
\caption{Testing strategies applied for all 12 factorial combinations of the building blocks.
For each combination one of the three testing strategies CTree/MOB/GUIDE is applied,
possibly after modifying the assessed variables as indicated (i.e., dichotomized or categorized).
\label{tab:combinations_appendix}}
\end{table}

\section{Significance level}\label{app:significance}

\begin{figure}[p!]
\setkeys{Gin}{width=\linewidth}
\includegraphics{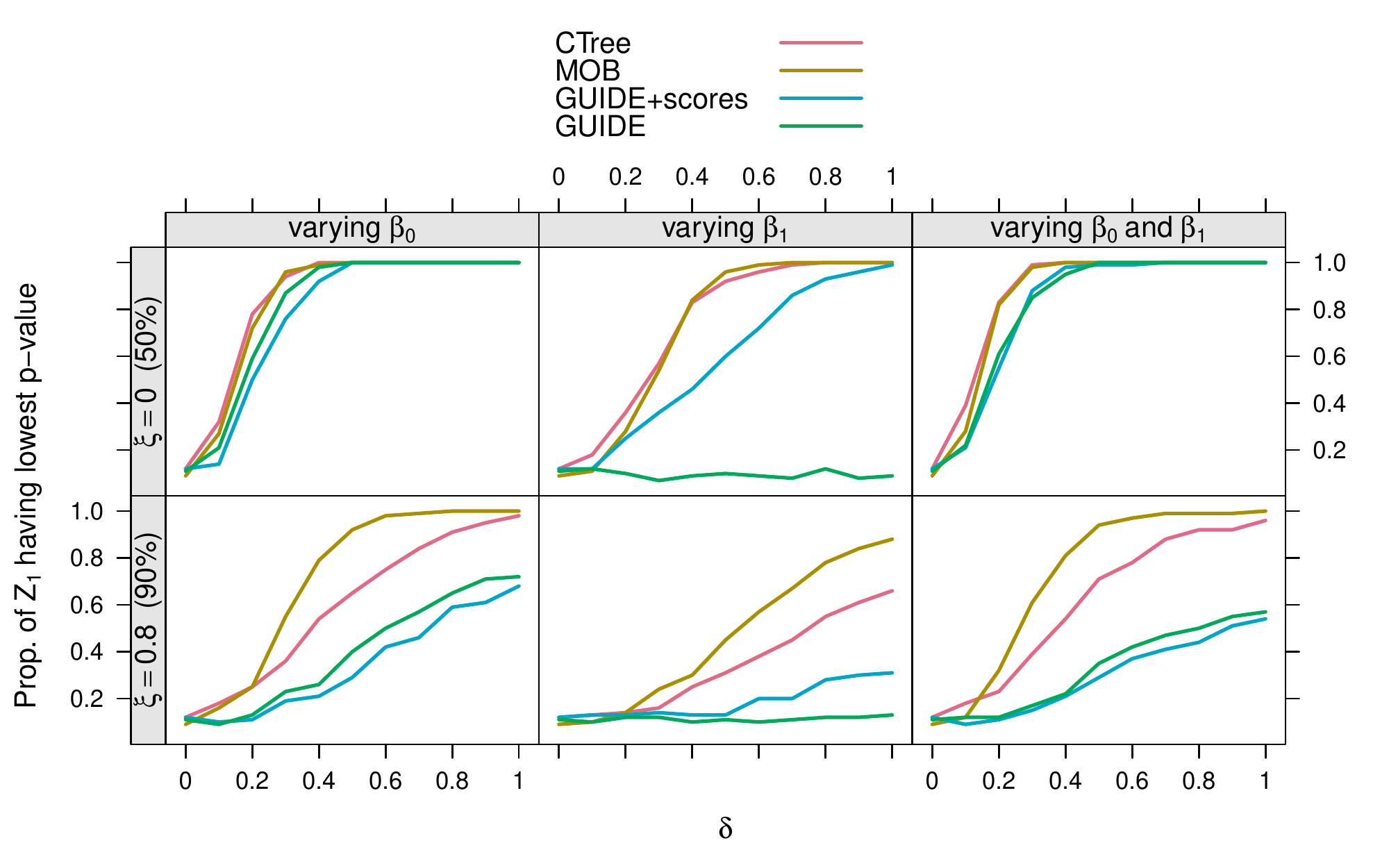}
\caption{\label{fig:scores_stump_wosign}
Proportion of replications where the true split variable $Z_1$ shows the smallest $p$-value, 
hence, where $Z_1$ is selected without any significance level in the ``stump'' scenario.
The testing strategies CTree, MOB, GUIDE, and GUIDE+scores are evaluated 
over increasing effect size $\delta$ with 100 replications of 250 observations per step. 
The true split point in $Z_1$ is either the median 0 (top) or the 90\% quantile 0.8 (bottom)
for either varying intercept $\beta_0$ (left), varying slope $\beta_1$
(middle), or both coefficients varying (right).}
\end{figure}

In Section~\ref{sec:scores} the investigated testing strategies are compared based
on the selection probability of $Z_1$, hence, the number of replications in which the true split 
variable $Z_1$ is detected with a $p$-value smaller than the predefined significance level 
$\alpha = 0.05$. This choice of measurement is due to the aim of investigating how well the 
tests perform as significance test.
However, ignoring this significance level yields the same conclusions which can be seen when 
comparing Figures~\ref{fig:scores_stump_wosign} and \ref{fig:scores_stump}, both being based 
on the exact same evaluation, but in Figure~\ref{fig:scores_stump_wosign} the proportion of 
replications in which the true splitting variable $Z_1$ is detected by showing the smallest $p$-value,
but not necessarily smaller than the significance level, is illustrated.

\section{Continuous ``stump'' scenario}
\label{app:contbeta}

\begin{figure}[t!]
\setkeys{Gin}{width=\linewidth}
\includegraphics{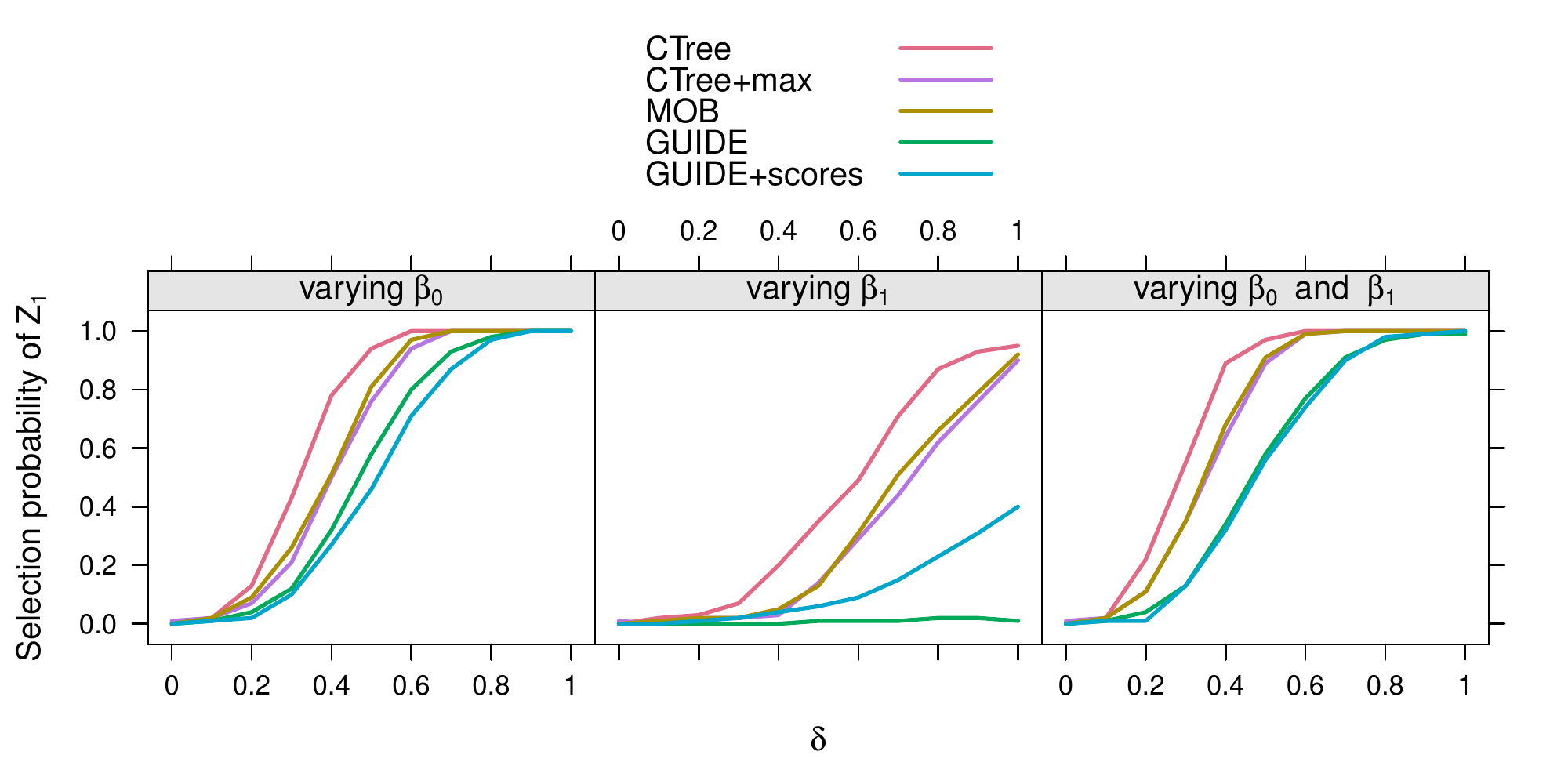}
\caption{\label{fig:stump_contbeta}
Selection probability of the true split variable $Z_1$ for testing strategies CTree, CTree+max,
MOB, GUIDE, and GUIDE+scores for continuous parameter functions $\beta_0(Z_1)$ and $\beta_1(Z_1)$.
Probabilities are estimated over increasing effect size $\delta$
with 100 replications of 250 observations per step. 
Each panel represents one variation for the coefficients: 
varying intercept $\beta_0$ (left), varying slope parameter $\beta_1$ (middle) and 
both coefficients varying (right).}
\end{figure}

In the simulation study presented in Sections~\ref{sec:simulation} and \ref{sec:results} the intercept
and slope parameters $\beta_0$ and $\beta_1$ are both either fixed or binary variables taking
either the positive or the negative value of the effect size $\delta$. Therefore, these parameters
are step functions for which the maximum selection employed in the testing strategy of MOB has shown 
to perform better than the linear selection in CTree. However, this changes in case of
monotonous functions where CTree is advantageous as it is constructed to detect monotonous effects.
To point out that the difference in performance of CTree and MOB depends on the type of effect
the same setting for which the results are presented in Figure~\ref{fig:scores_stump} is
evaluated again but this time with the varying parameter(s) changing continuously.
In particular, the parameters $\beta_0$ and $\beta_1$ are linear functions of the true
split variable $Z_1$:
\begin{align*}
\beta_{k-1}(Z_1) = \begin{cases}
-\delta \cdot (-1)^{k-1} \cdot Z_1\\
+\delta \cdot (-1)^{k-1} \cdot Z_1
\end{cases}
\end{align*}
for $k = 1,2=K$.
Additionally, a modified version of CTree employing a maximum selection, such as MOB, is evaluated 
(denoted by CTree+max). Looking at the selection probability illustrated in Figure~\ref{fig:stump_contbeta} 
it can be observed that in this setting the original version of CTree is ahead while CTree+max and MOB perform 
almost equally well. Hence, the comparison of Figures \ref{fig:scores_stump} and \ref{fig:stump_contbeta} 
points out that it clearly depends on the type of effect whether the maximum selection (MOB, CTree+max) 
or the linear selection (CTree) is to be preferred.

\section{Results for increasing effect size}
\label{app:increasingdelta}

\subsection{Dichotomization of residuals/scores}
\label{app:dichotomization}
To elaborate over an increasing effect size whether a dichotomization of the residuals/scores at zero 
leads to an improvement or a deterioration of performance, CTree and MOB are applied, once in their 
original version without dichotomization and once in an adapted version with a dichotomization of 
scores at zero (CTree+dich, MOB+dich). Moreover, they are compared to the adapted GUIDE version 
which includes all available scores (GUIDE+scores).
%Hence, the five strategies are obtained from the unifying framework by setting the switches 
%as listed in Table~\ref{tab:buildingblocks_appendix}.

Figure~\ref{fig:bin_stump} shows the effect of dichotomizing the score values over
different values for the true split point $\xi$, however, all four situations
lead to the same conclusions: Dichotomizing the score values decreases the selection probability
of the true split variable $Z_1$, and hence reduces the power of the testing strategy.

\subsection{Categorization of splitting variables}
\label{app:categorization}
To investigate the effect of categorizing the possible splitting variables CTree and MOB are
applied, once in their original version without a categorization and once in an adapted version
with a categorization of the possible split variables (CTree+cat and MOB+cat). 
Moreover, they are also compared to GUIDE+scores which includes all available scores.

\begin{figure}[h!]
\setkeys{Gin}{width=\linewidth}
\includegraphics{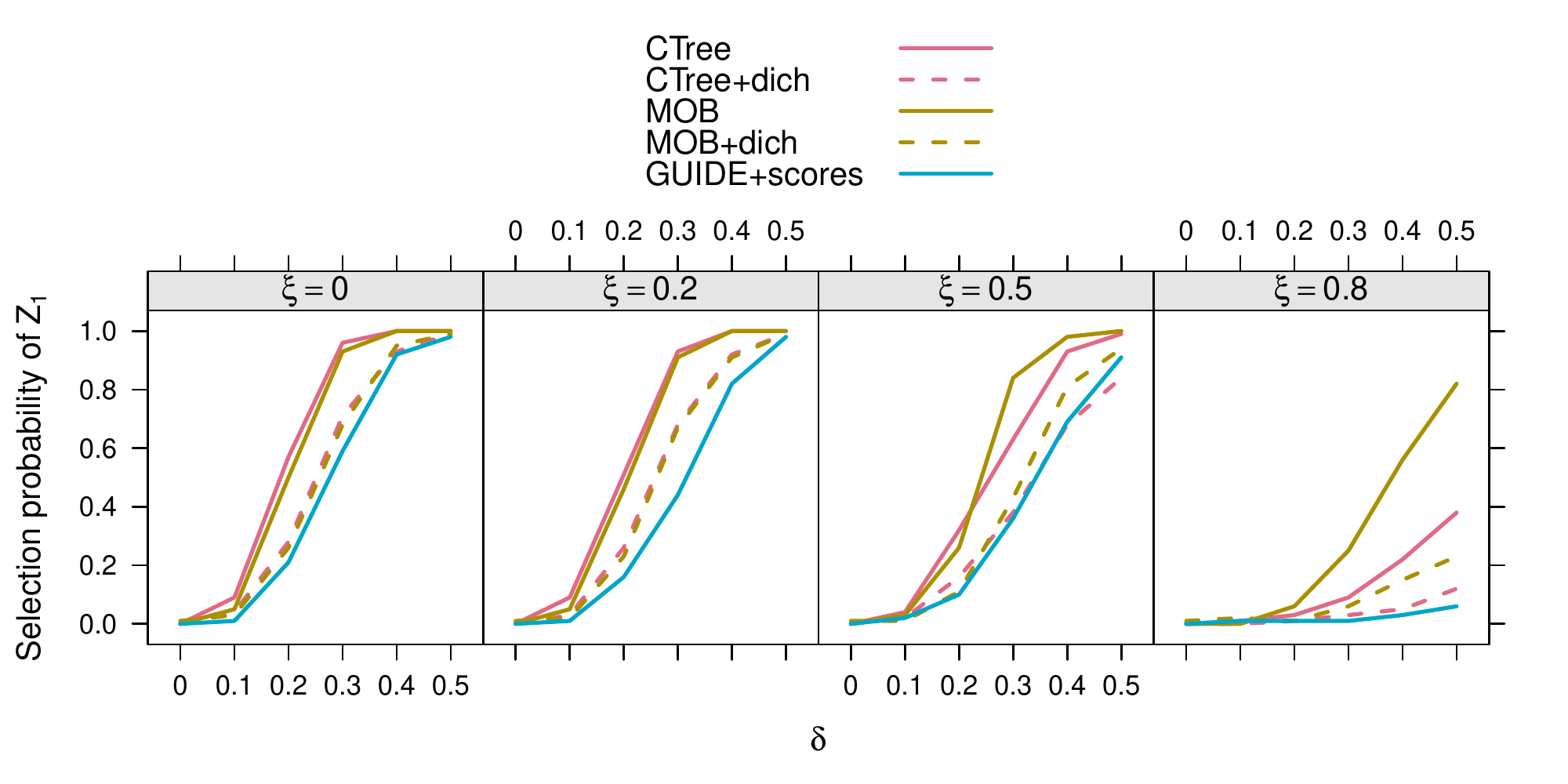}
\caption{\label{fig:bin_stump}
Selection probability of the true split variable $Z_1$ for
testing strategies CTree, CTree+dich, MOB, MOB+dich, and GUIDE+scores in the ``stump''
scenario with both parameters varying. CTree+dich and MOB+dich refer to modifications
of CTree and MOB applying a dichotomization of residuals/scores.
Probabilities are estimated over increasing effect size $\delta$
with 100 replications of 250 observations per step. 
Each panel represents a different value of the true split point $\xi$ (0, 0.2, 0.5 
and 0.8 from left to right corresponding to the 50\%, 60\%, 75\%, and 90\% quantile).
}

\setkeys{Gin}{width=\linewidth}
\includegraphics{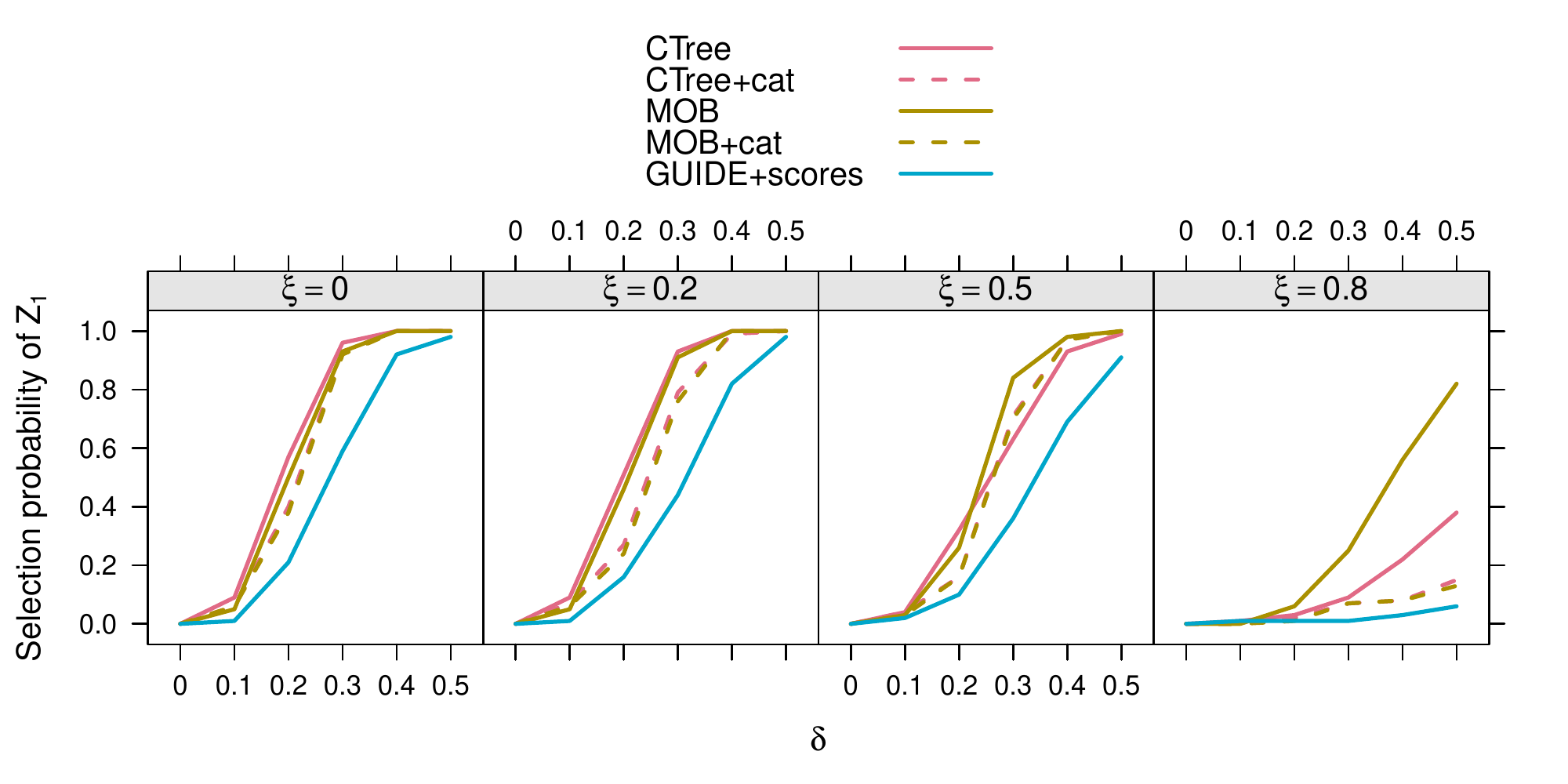}
\caption{\label{fig:cat_stump}
Selection probability of the true split variable $Z_1$ for
testing strategies CTree, CTree+cat, MOB, MOB+cat, and GUIDE+scores in the ``stump''
scenario with both parameters varying. 
Probabilities are estimated over increasing effect size $\delta$
with 100 replications of 250 observations per step. 
Each panel represents a different value of the true split point $\xi$ (0, 0.2, 0.5 
and 0.8 from left to right corresponding to the 50\%, 60\%, 75\%, and 90\% quantile).
}
\end{figure}

In Figure~\ref{fig:cat_stump} the impact of categorizing split variables on the performance 
is illustrated by the selection probability of $Z_1$ over increasing effect size and for four 
different values of the true split point $\xi$. For both, CTree and MOB, it can be stated that 
overall they perform better in their original form without categorization. Only if the true 
split point $\xi$ is close to the quartiles used as breaks for the categorizations both 
versions lead to a selection probability of $Z_1$ (e.g., for CTree and $\xi=0.5$)

Therefore, these results support the conclusions drawn in the main manuscript: Categorizing the values
of the split variable does not lead to any advantages unless the true split point corresponds (or is 
at least close) to one of the quartiles used for the categorization. In most situations it even causes
the power of the testing strategy to decrease.

\newpage

\clearpage

\end{appendix}

\end{document}